# Major axis kinematics of 15 early-type galaxies in the Fornax cluster. [*]

**Mauro D'Onofrio[1], Simone R. Zaggia[1], Giuseppe Longo[2], Nicola Caon[1], and Massimo Capaccioli[1,2]**

[1] Dipartimento di Astronomia, Universita di Padova, Vicolo dell'Osservatorio 5, I-35122 -- Padova -- Italy
[2] Osservatorio Astronomico di Capodimonte, Via Moiariello 16, I-80131 -- Napoli -- Italy

**Abstract.** Major axis rotation curves and velocity dispersions profiles, extending out to about one effective radius, are presented for 15 ellipticals and S0's out of the photometric sample of Fornax cluster galaxies studied by Caon et al. (1994). A brief description of the spectroscopic and photometric characteristics of each galaxy is provided, together with a comparison with previous studies. Six of the nine E's are possibly misclassified S0's or ellipticals harboring a disk-like component. Two galaxies (NGC 1399 and 1404) show a hint of counter-rotation.

Evidence is given that the *bright* and the *ordinary* families of early-type galaxies, first introduced by Capaccioli et al. (1992), look distinct also in term of the anisotropy parameter $(V_{\rm m}/\sigma)^*$.

**Key words:** Galaxies: clusters: Fornax -- galaxies: elliptical and lenticular -- galaxies: kinematics and dynamics

## 1. Introduction

Prompted by the need of adopting standard methodologies, procedures, and strategies for the study of the global properties of early-type galaxies (cf. Capaccioli et al. 1993a), Caon et al. (1990, 1994; hereafter C[2]D) have recently carried on a program of accurate surface photometry of two luminosity- and volume-limited samples of E and S0 galaxies belonging to the Virgo and Fornax clusters. The C[2]D data-set has been already utilized for a study of the photometric properties of faint and bright E and S0 galaxies in a series of papers. Capaccioli et al. (1992, 1993b), for instance, have analyzed the distribution of C[2]D galaxies in the plane defined by the global parameters $r_{\rm e}$ (effective radius) and $\mu_{\rm e}$ (corresponding surface brightness). The main finding is that, at variance with previous belief (e.g. Kormendy 1977, Djorgovski and Davis 1987), this morphological class is heterogeneous as for the global parameters, and it encompasses two photometric families, named *bright* and *ordinary*, which appear distinct also in relation to the environmental characteristics and to the internal kinematics, color gradient, radio and X-ray emission, dust and hydrogen content.

Moreover, by looking at the C[2]D data-set, Caon et al. (1993) and D'Onofrio et al. (1994) have found that the light distribution of early-type galaxies is shaped according to the total luminosity, i.e. the light profiles are better represented by a $r^{1/m}$ formula than by the universally adopted $r^{1/4}$ law (de Vaucouleurs 1948), with $m$ increasing monotonically from dwarfs to giants.

Along the same vein Jrgensen et al. (1992) and Jrgensen & Franx (1994) have started a systematic analysis of the photometric and geometric properties of the early-type galaxies which are members of the Coma cluster, finding that faint E and S0 galaxies form a continuous class of objects, different from the giants, but also from the dwarfs.

The observed similarities in the photometric properties of normal ellipticals and S0's raises the question if they share also similar dynamical properties (see Capaccioli & Longo 1994 for a review). It is known that, while most normal E galaxies are rotationally flattened, the brightest ellipticals are supported by the anisotropic velocity dispersion (Bertola & Capaccioli 1975, Illingworth 1977, Binney 1978). The anisotropy parameter $(V_{\rm m}/\sigma)^*$ --- where $V_{\rm m}$ is the maximum rotational velocity, $\sigma$ is some average velocity dispersion, and their ratio is normalized to the values pertinent to an oblate rotator (Davies et al. 1983) --- correlates with the effective radius $r_e$ (Capaccioli et al. 1993b) and with the isophotal shape parameter $a_4$ (Bender et al. 1989).

The low surface brightness dwarf ellipsoidal systems are mostly supported by the anisotropic velocity dispersion (Bender & Nieto 1990), indicating that the kinematical properties do not vary smoothly from the dwarf to the giant types, as suggested by the light profiles. However, it is presently unclear how much the amount of pressure-support is correlated with the morphology for galaxies of different luminosities.

A more complete understanding of the properties of the early-type galaxies awaits larger and more homogeneous data-sets. At present, the only sample of spectroscopic data fulfilling part of these requirements is that of the 7 Samurai (Davies et al. 1987) which, however, does not provide extended rotation curves and velocity dispersion profiles. Sparse data can be found in quite a few papers, as those dealing with the tridimensional structure of ellipticals and the peculiar motions; for updated compilations see Busarello et al. (1989, 1994). It is for these reasons that we decided to embark in a program of detailed spectroscopic observations for galaxies in the $C^2D$ sample.

This paper reports on the major axis rotation curves (*RC*) and velocity dispersion (*VD*) profiles for 15 early-type galaxies members of the Fornax cluster. The observational material and the most important steps of the data reduction are described in Sections 2 and 3. We comment on the individual galaxies in Sect. 4, compare our results with the literature in Sect. 5, and discuss the correlation of the kinematical and photometrical parameters in Sect. 6.

## 2. Observations

Spectra of 15 early-type galaxies of the Fornax cluster were collected in November 1991 with the B&C spectrograph attached at the Cassegrain focus of the ESO 1.5 m telescope. The objects, selected from Ferguson's (1989; hereafter F89) catalogue of the Fornax cluster, are E's and S0's in common with the $C^2D$ sample, with the exception of IC 1963. This first contribution, in a long-range project aimed at providing kinematics data for all the galaxies of the $C^2D$ dataset, deals with 62% of the Fornax population of E's and S0's (according to the membership list of F89).

The target list is presented in Table 1; columns from 1 to 6 give the names of the galaxies, the morphological types according to F89 and de Vaucouleurs et al. (1991; hereafter RC3) catalogue, the total luminosities $B_T$ and the position angles (P.A.) of the major axes from $C^2D$, and the epochs of the spectra.

**Table 1.** Target list

| Galaxy Name | Morph. Type | RC3 | $B_T$ [mag] | P.A. [deg] | Date (1991) |
|---|---|---|---|---|---|
| IC 1963 | S0$_1$(9) | .L..../ | 13.00 | 84 | 12/11 |
| NGC 1316 | S0$_3$ pec | PLXS0P. | 9.23 | 50 | 11/11 |
| NGC 1336 | E4 | .LA.-.. | 13.27 | 20 | 16/11 |
| NGC 1339 | E4 | .E+..P* | 12.70 | 175 | 12/11 |
| NGC 1351 | E5 | .LA.-P* | 12.30 | 141 | 15/11 |
| NGC 1374 | E0 | .E..... | 11.94 | 119 | 15/11 |
| NGC 1375 | S0(cross) | .LX.0*/ | 13.21 | 88 | 16/11 |
| NGC 1379 | E0 | .E..... | 12.03 | 7 | 11/11 |
| NGC 1380 | S0/a | .LA.... | 10.82 | 7 | 14/11 |
| NGC 1380A | S0$_2$(9) | .L..0*/ | 13.29 | 177 | 12/11 |
| NGC 1381 | S0(9) | .LA..*/ | 12.32 | 139 | 15/11 |
| NGC 1399 | E0 | .E.1.P. | 10.01 | 112 | 11/11 |
| NGC 1404 | E2 | .E.1... | 10.93 | 159 | 14/11 |
| NGC 1419 | E0 | .E...P* | 13.56 | 65 | 16/11 |
| NGC 1427 | E4 | .E+.... | 11.82 | 79 | 14/11 |

The detector was the ESO CCD #27 Ford chip, which has a very good cosmetics and a readout noise of 3.1 e$^-$. To limit the readout time, the CCD was windowed to 2048 × 500 pixels. This subframe is large enough to contain the dark level of the science exposures in the parts not covered by the spectrum. We used grating #26 (1200 mm$^{-1}$) giving a dispersion of 1.0 A pix$^{-1}$ over a range of 2000A centered at 5900A. The spectrograph slit was chosen to be 220 $\mu m$ wide (2″ on the sky) and 2″.9 long (the maximum value allowed). The scale in the direction perpendicular to the dispersion is 0″.68 pix$^{-1}$.

Each galaxy was observed twice, with a typical exposure time of one hour per spectrum. The slit was centered on the galaxy nucleus and aligned with the photometric major axis (Table 1). We had to reject one of the two spectra of NGC 1316 and of

**Fig. 1.** Comparison of the light profiles along the major axis of NGC 1375 from high precision surface photometry ($C^2D$; solid line) and from one spectrum of the galaxy (open circles).

NGC 1379 due to bad tracking of the telescope. Template spectra of KIII standard stars were obtained each night, together with spectra of the dawn sky-light (in order to check the flatness of the slit profile), and the usual series of bias, dark, and flat-field exposures.

## 3. Data reduction

The first part of the data reduction was almost standard, except for one problem. We realized that the spectra were misaligned with respect to the CCD columns, with a difference of a maximum of 6 pixels ($\sim 4''$) between the two extremes of the spectra. This defect which, if overlooked, would have affected both the *RC* and the *VD* measurements, was corrected as follows. For each CCD column (i.e. in the direction perpendicular to the dispersion), a gaussian curve was fitted to the light profile of the object (galaxy or template), and the position of the maximum measured. Then, by correlating the column number with the position of the maximum, we derived the angle which was used to rigidly rotate the spectrum.

Each He-Ar comparison spectrum was corrected by the same angle applied to the scientific exposure before and after which they had been taken. We found in fact that the tilt of the spectra varied in an unpredictable way with the position of the telescope and the rotation of the spectrograph. Using the two spectra for each galaxy, we verified that the rotation angles of the spectra were the same.

Cosmic-ray particle events were interactively removed by means of a bidimensional fit in a window three times larger than the area containing the cosmic spike. Then, we determined the wavelength calibration for each spectrum with an accuracy of $\sim 0.1$ A.

The sky spectrum was subtracted from the science exposure by using that part where the contribution of the galaxy light was supposedly negligible. In order to check the accuracy of this operation, an average light profile was produced for each galaxy spectrum by co-adding all columns (cf. Longo et al. 1994) and compared with the photometry of the major axis light profile from $C^2D$. Despite the differences in the photometric bands, the two profiles match each other very well (see the example of NGC 1375 in Fig. 1).

Finally, the calibrated spectrum was rebinned in $\log(\lambda)$ units, the continuum was subtracted row-by-row, and the result processed through a low and a high-bandpass filter.

The spectra were analyzed by the Fourier Correlation Quotient (FCQ) method (Bender 1990), which is based on the deconvolution of the correlation peak of the template-galaxy correlation function with the peak of the autocorrelation function of the template. This allows to recover the broadening-function of the spectrum --- which is not analyzed here to investigate the velocity distribution along the line of sight, since this would demand a much higher signal--to--noise (S/N) ratio.

The internal consistency of our results is certified by the agreement between the solutions for the two spectra of each galaxy. As a test we repeated the FCQ analysis using different template stars; as usual, we did not note any significant difference arising from the mismatch of the spectral types.

The final *RC* and *VD* profiles, presented in graphical form in Appendix, were extracted from the sum of the two spectra (when available), in order to increase the S/N ratio. The rotation curves were first folded around the luminosity center of the galaxy and then best matched through shifts in radial velocity and spatial position, in order to find the kinematical center under the assumption of *maximum kinematical symmetry*. Points and errors in the figures come from the weighted mean of the two sides of the spectrum and are rebinned in radius using an increasing step according to the local S/N.

The basic data for the present sample of galaxies are collected in Table 2, which reports, from column 1 to 9, the galaxy name, the radius of the maximum extension of the spectrum $r_{lim}$ and the corresponding surface brightness $\mu_{lim}$, the ratio $r_{lim}/a_e$ where $a_e$ is the length of the effective semimajor axis of the galaxies, the central velocity dispersion $\sigma_0$ and its adopted error, the maximum observed rotation velocity $V_m$ and the estimated error, the maximum ellipticity of the galaxy, the recession velocity with its error, and the shift $\Delta r$, giving the amount and direction of the displacement of the kinematical center with respect to the photometric center.

## 4. Comments on individual objects

Here we briefly describe and discuss the photometric, geometric, and kinematical properties of the sampled galaxies. The *RC* and the *VD* profiles along the major axis are shown in the Appendix. The corrisponding tables can be collected via an anonymous ftp at the CDS.

**Table 2.** Kinematical parameters

| Galaxy | $r_{lim}$ | $\mu_{lim}$ | $r_{lim}/a_e$ | $\sigma_0$ | $V_m$ | $cz$ | $\varepsilon_m$ | $\Delta r$ |
|---|---|---|---|---|---|---|---|---|
| IC 1963 | 45 | --- | 2.0: | 80± 3 | 120±10 | 1619± 6 | --- | 0.1 |
| NGC 1316 | 52 | 21.0 | 0.4 | 260±10 | 135±15 | 1789±11 | 0.43 | 3.5 |
| NGC 1336 | 12 | 22.2 | 0.4 | 115± 6 | 10±10 | 1421± 8 | 0.36 | 0.0 |
| NGC 1339 | 26 | 22.6 | 1.7 | 190± 6 | 145±10 | 1367± 8 | 0.31 | 0.0 |
| NGC 1351 | 19 | 21.8 | 0.6 | 155± 6 | 165±20 | 1511± 8 | 0.38 | −0.7 |
| NGC 1374 | 19 | 21.7 | 0.7 | 225±10 | 55±20 | 1332±11 | 0.12 | −0.4 |
| NGC 1375 | 21 | 21.8 | 0.9 | 80± 4 | 80±20 | 740± 7 | 0.61 | −1.0 |
| NGC 1379 | 16 | 21.3 | 0.7 | 140±10 | 45±10 | 1380±11 | 0.04 | 0.3 |
| NGC 1380 | 65 | 22.1 | 1.2 | 260±12 | 250±50 | 1877±13 | 0.57 | 0.0 |
| NGC 1380A | 34 | 22.1 | 1.1 | 80± 5 | 80±15 | 1561± 7 | 0.83 | 3.0 |
| NGC 1381 | 52 | 22.5 | 2.2 | 165± 6 | 165±10 | 1724± 8 | 0.76 | −1.1 |
| NGC 1399 | 46 | 22.2 | 0.3 | 420±27 | 80±30 | 1465±27 | 0.13 | −0.3 |
| NGC 1404 | 31 | 21.7 | 1.2 | 260±11 | 130±30 | 1912±12 | 0.15 | −0.1 |
| NGC 1419 | 5 | 20.5 | 0.5 | 125± 8 | 10±10 | 1560± 9 | 0.18 | 0.0 |
| NGC 1427 | 31 | 22.2 | 0.8 | 180± 5 | 45±10 | 1416± 7 | 0.34 | 0.0 |

Notes to the table: Col. 2: $r_{lim}$ in arcsec. Col. 3: B-band surface brightness in mag·arcsec$^{-2}$. Cols. 5,6,7: Velocity dispersion, maximum rotational velocity and recession velocity in km s$^{-1}$. Col. 8: maximum observed ellipticity. Col. 9: Shift $\Delta r$ in arcsec: + in the direction of the P.A.

**IC 1963:** No surface photometry is available for this edge-on galaxy; the value of $r_{lim}/a_e$ listed in Table 2 rests on the measure of $r_e = 11.3''$ by F89[1] and on the flattening quoted in RC3 ($b/a = 0.25$). The $VD$ profile is at the limit of our resolution. The $RC$ curve is typical of a low mass S0; it flattens up at $\sim 20'' \equiv 1\ a_e$.

**NGC 1316:** Well known peculiar S0 (or D galaxy) with dust patches, particularly evident along the minor axis (Schweizer 1980, 1981). The light profiles along the main axes follow approximately the $r^{1/4}$ law. Ellipticity ranges between 0.3 and 0.4 at all $r \gtrsim 5''$. The Fourier coefficients (hereafter $FC$'s) describing the residuals from the ellipse fitting of the isophotes show a complex behaviour. The measured central velocity dispersion ($\sigma_0 = 260$ km s$^{-1}$; Table 2) is somewhat lower than expected for a galaxy of this luminosity on the ground of the Faber-Jackson relation ($\sigma_{FJ} \approx 400$ km s$^{-1}$). The flattening is almost entirely supported by rotation according to the position of the object in the $V_m/\sigma - \varepsilon$ diagram (Fig. 6). The ratio $(V/\sigma)$ increases with radius from $a > 5''$. Our spectrum shows H$\alpha$, the [NII] satellites and the [SII] doublet in emission, extending out to $r \simeq 5''$ SE of the nucleus and to $r \simeq 30''$ NW.

This object present the maximum shift $\Delta r$ of the kinematical and optical centers.

**NGC 1336:** Faint galaxy with low $VD$ within the observed range ($r \lesssim 10''$). Here the isophotes are almost round ($\varepsilon < 0.1$), the $(V/\sigma)$ ratio is constant, and there is (possibly) a large isophotal twist. The Fourier coefficients suggest the presence of an inner component, either a face-on disk or more likely a bar, which might account for the observed low rotation.

**NGC 1339:** The galaxy is classified E4 but it is very likely an S0. Both the major and minor axis light profiles show systematic deviations from the $r^{1/4}$ law in the outer regions which are typical of the presence of a disk; also the $a_4$ $FC$ is moderately but consistently positive over the whole range covered by C$^2$D photometry. The gradient in the first $4''$ is $\sim 25$ km s$^{-1}$ per arcsec. The ellipticity increases from 0.1 at the center to 0.3 at $\sim 8''$, and the $(V/\sigma)$ ratio increases from 0.2 to 1.5, suggesting a transition between an inner bulge and an outer disk.

**NGC 1351:** Low $VD$ and large gradient in the inner parts of the $RC$ ($\sim 35$ km s$^{-1}$ per arcsec in $2''$). The isophotal analysis suggests the presence of an inner component stretching along the major axis. However, the morphology of the rotation and velocity dispersion profiles suggests instead an inner bulge (to $r \sim 10''$) and an outer disk. Also, the classification of the galaxy as lenticular (RC3) is likely due to another elongated component (disk ?) at $r > 0'.5$.

**NGC 1374:** Large central gradient of the $VD$ ($\sim 90$ km s$^{-1}$ per arcsec in $4''$). The rotation velocity is low but with a trend typical of a disk galaxy. The flattening is also low and the light profiles follow quite well the $r^{1/4}$ law. This object could be a face-on S0.

**NGC 1375:** Small gradient in the $RC$ and low $VD$. The bulge of this edge-on S0 is very boxy inside $r \sim 15''$, a shape due to the superposition of a strong disk with a ring-like structure seen in absorption, according to Lorenz et al. (1993). The ellipticity increases from 0 to 0.6 in $20''$.

---

[1] Individual effective radii in F89 are consistent, within $\sim 20\%$ s.e., with those in C$^2$D.

<u>NGC 1379:</u> Nearly constant $VD$ out to $r \sim 7''$ and low rotation velocity. The ellipticity is very low and the isophotes moderately disky. This galaxy may be a face-on S0, in which the disk component starts at $r = 5''$. The morphology of the $RC$ and of the light profiles support this hypothesis.

<u>NGC 1380:</u> S0/a galaxy with probably a dusty nucleus. Inside $r = 5''$ there is a strong variation of the ellipticity, position angle, and $FC$'s. The gradient in the $RC$ and the $VD$ is high: $\Delta V = 90$ km s$^{-1}$ in $5''$ and $\Delta\sigma = 100$ km s$^{-1}$ in $20''$. The $(V/\sigma)$ ratio increases outside $20''$ where the disk dominates. A large peak in $a_4$ is present at $r \sim 1'$, possibly a ring. The galaxy is close to the line of the isotropic oblate rotators in the $V_m/\sigma - \varepsilon$ diagram (Fig. 6).

<u>NGC 1380A:</u> Faint edge-on S0. The $RC$ extends out to $0'.5$ and has a small gradient. The $VD$ is low and nearly constant, the ellipticity is very high and the rotation velocity is flat from $r \simeq 15''$. The $(V/\sigma)$ ratio is approximately constant inside $10''$, then increases with $\varepsilon$. The maximum of $a_4$ ($\sim 18$) measured at $r \sim 17''$ does not coincide with the change of slope observed in the major axis light profile at $r \sim 30''$. The latter has a trend typical of a disk component seen nearly edge-on. The minor axis light profile follows the $r^{1/4}$ law.

For this galaxy we found an high displacement between the kinematical and the optical centers. However, in the first 6 arcsec the $RC$ is almost zero and $\Delta r$ is quite uncertain.

<u>NGC 1381:</u> Disk-dominated S0 with large gradient in the $VD$ inside $10''$. A small core or a kinematically distinct component could be present inside $r \sim 6''$, judging from the local minimum in the $RC$ and $VD$ profiles. The $(V/\sigma)$ ratio increases monotonically with $\varepsilon$. The galaxy is close to the line of the isotropic prolate rotators (Fig. 6). A large positive feature in the $FC$'s is present at $r \sim 30''$ (ring ?), but there are no features in the $RC$ at this radius.

<u>NGC 1399:</u> Bright E1 galaxy. The rotation velocity is very low, but possibly increases from $r = 30''$ on. The $VD$ has a large gradient in the first $10''$ ($\sim 13$ km s$^{-1}$ arcsec$^{-1}$) and the $(V/\sigma)$ ratio is low and nearly constant. At the center there is a hint of counter-rotation, which may suggest the occurrence of a kinematically decoupled core or may be due to strong triaxiality.

<u>NGC 1404:</u> The $VD$ is high and nearly constant up to $20''$, but in the core ($r < 3''$) possibly decreases by $\sim 20$ km s$^{-1}$. In the inner $15''$ the $RC$ increases rapidly (the gradient is $\sim 10$ km s$^{-1}$ arcsec$^{-1}$) then it flattens up (the following decrease is likely unreal). The ellipticity is always lower than 0.16, the major axis twists by $\sim 30°$, and there are two peaks in the $a_4$ profile. The galaxy is near to the lines of the isotropic rotators (Fig. 6), thus suggesting that it could be a face-on S0. The $RC$ has a hint of counter-rotation in the very center, but the feature is small in size and amplitude. Moreover, the analysis of the (low S/N) broadening function does not reveal any evidence of asymmetries.

<u>NGC 1419:</u> Low luminosity E0 galaxy. The object has a large isophotal twist in the inner $20''$, where $\varepsilon < 0.1$. The ellipticity increases up to 0.2 at $60''$. There is almost no rotation and the $VD$ is low and nearly constant.

<u>NGC 1427:</u> Galaxy of intermediate intrinsic luminosity. The $VD$ is quite high and nearly constant out to $20''$. The $RC$ has a peak at $4''$, and there is almost no rotation outside.

The $FC$'s give marginal evidence for the presence of an inner disk. In the $V_m/\sigma - \varepsilon$ diagram the galaxy is far from the isotropic oblate rotators.

**Fig. 2.** NGC 1316: comparison of our major axis kinematical measurements (solid line) with Bosma's et al. (1985; open circles) and Jenkins & Scheuer's (1980; open squares).

## 5. Comparison with literature data

Only few of our galaxies are in common with other kinematical studies. Fig. 2 shows that our results on NGC 1316 compare well with those of Bosma et al. (1985), but not with Jenkins & Scheuer's (1980). Bosma et al. found a lower value of the central dispersion of approximately 30 km s$^{-1}$, but the general trend of the $RC$ and $VD$ profiles are the same. We do not observe the jump in the $RC$ at the position of the first ripple ($\sim 25''$), but rather a rapid increase of the rotation velocity followed by a progressive flattening. In the $VD$ profile we have instead a jump at the same position, where the velocity dispersion increases by $\sim 30 \div 40$ km s$^{-1}$. The position angle of the slit along the major axis was different by 10 degree in the two cases.

Jenkins & Scheuer (1980) showed a larger value of the central velocity dispersion (of $\sim 20$ km s$^{-1}$) and found also higher values at $\sim 10''$ both in the $RC$ and in the $VD$ profiles (respectively of 130 km s$^{-1}$ and of 300 km s$^{-1}$).

We have three galaxies in common with Franx et al. (1989): NGC 1379, NGC 1399, and NGC 1404. The agreement with our results is quite good, with only a small discrepancy in the $VD$ for NGC 1379 and NGC 1399 (Fig. 3 $a$--$c$). As noted above, our data on NGC 1404 show an asymmetry at the center of the $RC$, suggesting some sort of kinematical decoupling. We have verified the presence of such an asymmetry using an EMMI-NTT spectrum of NGC 1404, kindly taken and reduced for us by W.W. Zeilinger. The velocity resolution and the S/N ratio are higher than in our 1.5 m telescope spectrum, and confirm that inside $2''$ there is a peculiar velocity distribution, possibly an inversion of the velocity gradient. However, the small amplitude ($\sim 10$ km s$^{-1}$) makes this feature still uncertain.

We have compared the *RC* of NGC 1380 and NGC 1381 with Dressler & Sandage's (1983) (Fig. 4). For NGC 1380 the *RC* runs higher by $\sim 40 - 50$ km s$^{-1}$ at $r > 10''$, while for NGC 1381 our *RC* is somewhat lower (of about 20--25 km s$^{-1}$) in the inner $20''$.

Ten galaxies of our sample are in common with the 7 Samurai, and four respectively with the catalogues of central velocity dispersion by Whitmore et al. (1985) and Davoust et al. (1985). The largest discrepancies with our data are found for the central velocity dispersion of NGC 1399 and NGC 1404, that we measure higher by $\sim 100$ and $\sim 50$ km s$^{-1}$ respectively (Fig. 5). Note the trend in the data of the 7 Samurai of measuring progressively lower velocity dispersions with respect to our results. The most discrepant galaxy is NGC 1399 whose value of $\sigma_0$ is confirmed by Stiavelli et al. (1993); their *VD* profile and the steep gradient match well our data (Fig. 3).

## 6. Morphology versus kinematics

The $V_m/\sigma - \varepsilon$ diagram (Binney 1978) gives some clues on the intrinsic shape of galaxies. Figure 6 shows this diagram for our 15 galaxies: circles and triangles represent galaxies with boxy and disky isophotes respectively, while squares are for undefined objects and pentagon's for inner--boxy and outer--disky objects. Here we used as variables the maximum observed rotation velocity, the central velocity dispersion and the maximum ellipticity. It is interesting to note that, among the brightest galaxies, NGC 1316 and NGC 1404 lie near the lines of the isotropic rotators (solid line for the oblate case, dashed for the prolate). This is no surprise for NGC 1316, classified as a peculiar S0, but it is for the E2 NGC 1404.

**Fig. 3.** From top to bottom NGC 1379, NGC 1399 and NGC 1404: comparison of our major axis kinematical measurements (solid line) with Franx's et al. (1989; open circles) and, for NGC 1399 only, with Stiavelli et al. (1993; open triangles).

Another important dynamical parameter is $(V_m/\sigma)^*$, i.e. the ratio of the maximum rotational velocity to the velocity dispersion, both measured in units of the values expected for an oblate rotator (Davies et al. 1983). It is defined here as $(V_m/\sigma)^* = (V_m/\sigma_0)/\sqrt{\varepsilon(1-\varepsilon)}$, using $\sigma_0$ instead of the mean value inside $r_e$. The plot of $(V_m/\sigma)^*$ versus the total luminosity for our galaxies is shown in Fig. 7 (filled squares). We also plotted the averaged data for the early-type galaxies appearing in the kinematical catalogue of Busarello et al. (1994), indicating ellipticals and S0's respectively as open circles and squares. Asterisks are the data for the faint ellipticals from Bender & Nieto (1990), and triangles are the bulges of spirals from Kormendy & Illingworth (1982). Despite the large scatter, mainly due to errors in the data, it is apparent that there are almost no galaxies of intermediate luminosity ($-18.2 \lesssim M_B \lesssim -20.2$, assuming a distance of 19.6 Mpc for the Fornax cluster; Capaccioli et al. in preparation) dominated by anisotropic velocity dispersion ($\log(V_m/\sigma) \lesssim -0.25$). Ellipticals, S0's, and bulges are well mixed in this range of luminosities.

**Fig. 4.** Rotation curves of NGC 1380 and NGC 1381. Open circles are data from Dressler & Sandage (1983).

Finally, we show in Fig. 8 the correlations between the anisotropy parameter, the mean surface brightness, and the effective radius for a sample of Fornax and Virgo galaxies which possess accurate kinematical[2] and photometric data. Galaxies of the *bright* and the *ordinary* families, as defined in Capaccioli et al. (1992), are represented with filled and open symbols respectively.

**Fig. 5.** Comparison of our $\sigma_0$ with the data from: the 7 Samurai, Whitmore et al. (1985, WMT) and Davoust et al. (1985, DPV).

We see that, although with large scatter, there is a rough correlation among the variables. As found by Davies et al. (1983), the higher surface brightness galaxies have higher anisotropy parameters and smaller effective radii. They have probably experienced more dissipation phenomena during their formation. On the other hand, the galaxies members of the *bright* family, which are likely the result of merging processes, have lower $V_{\rm m}/\sigma$ ratios and higher effective radii.

---

[2] Virgo data are from the catalogue of Busarello et al. (1994).

**Fig. 6.** The $V_{\rm m}/\sigma - \varepsilon$ diagram for the Fornax galaxies of our sample. The solid curve represents the isotropic oblate ellipsoids with isotropic velocity dispersion. The dashed curve is for isotropic prolate rotators.

**Fig. 7.** The anisotropy parameter $\log(V_{\rm m}/\sigma)^*$ against the absolute magnitude $M_{\rm B}$ ($H_0 = 70$ km s$^{-1}$Mpc$^{-1}$). Symbols are coded in the text. The dashed lines frame the interval $\pm 0.25$ in $\log(V_{\rm m}/\sigma)^*$, where oblate rotators are expected to fall.

**Fig. 8.** *Top left:* Correlation between the anisotropy parameter $\log(V_{\rm m}/\sigma)^*$ and the logarithm of the effective radius $R_{\rm e}$ in kpc. *Top right:* $\log(V_{\rm m}/\sigma)^*$ versus the mean surface brightness $\langle \mu \rangle_{\rm e}$ in the B-band. *Bottom left:* The $\langle \mu \rangle_{\rm e} - \log r_{\rm e}$ relation. Filled and open symbols represent galaxies of the *bright* and the *ordinary* families respectively.

Our analysis of the $\log(V_{\rm m}/\sigma)^* - \langle \mu \rangle_{\rm e}$ relation (Wyse & Jones 1984) has shown that the residuals do not correlate clearly with the ellipticities or with the effective radii. However, this result is based on few objects and should be considered only as indicative. Probably a number of concomitant causes contribute to the scatter in the diagrams, such as inclination effects, the different amount of anisotropy, and the random errors in the velocities and in the effective radii.

*Acknowledgements.* We warmly thank Romano Corradi for providing us with further observations for this program in January 1992, and Werner Zeilinger for his spectral analysis of the galaxy NGC 1404 observed at NTT.

## A. Tables and Figures

Tables A1 to A15 (available in electronic form) and the following figures present the major axis rotation curves and velocity dispersion measurements of the sampled galaxies. In each table $r$ is the distance from the dinamical center along the major axis in arcsec, $V(r)$ is the rotation velocity with its error $\Delta V$, $\sigma(r)$ is the velocity dispersion and $\Delta\sigma$ its error.

In the figures solid dots represent our measurements, and the open circles are those values of the velocity dispersion near the velocity resolution limit.

Fig. 1

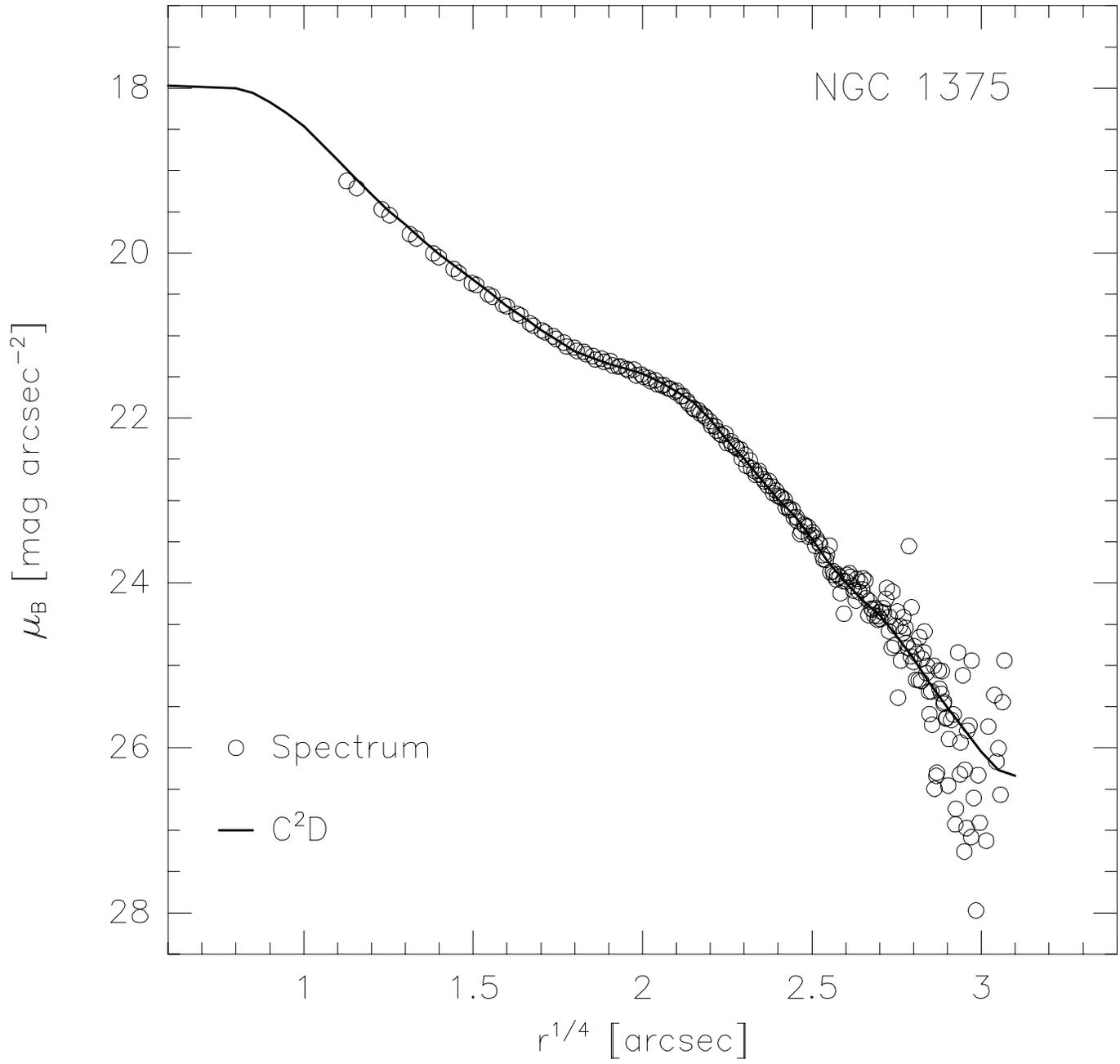

Fig. 2

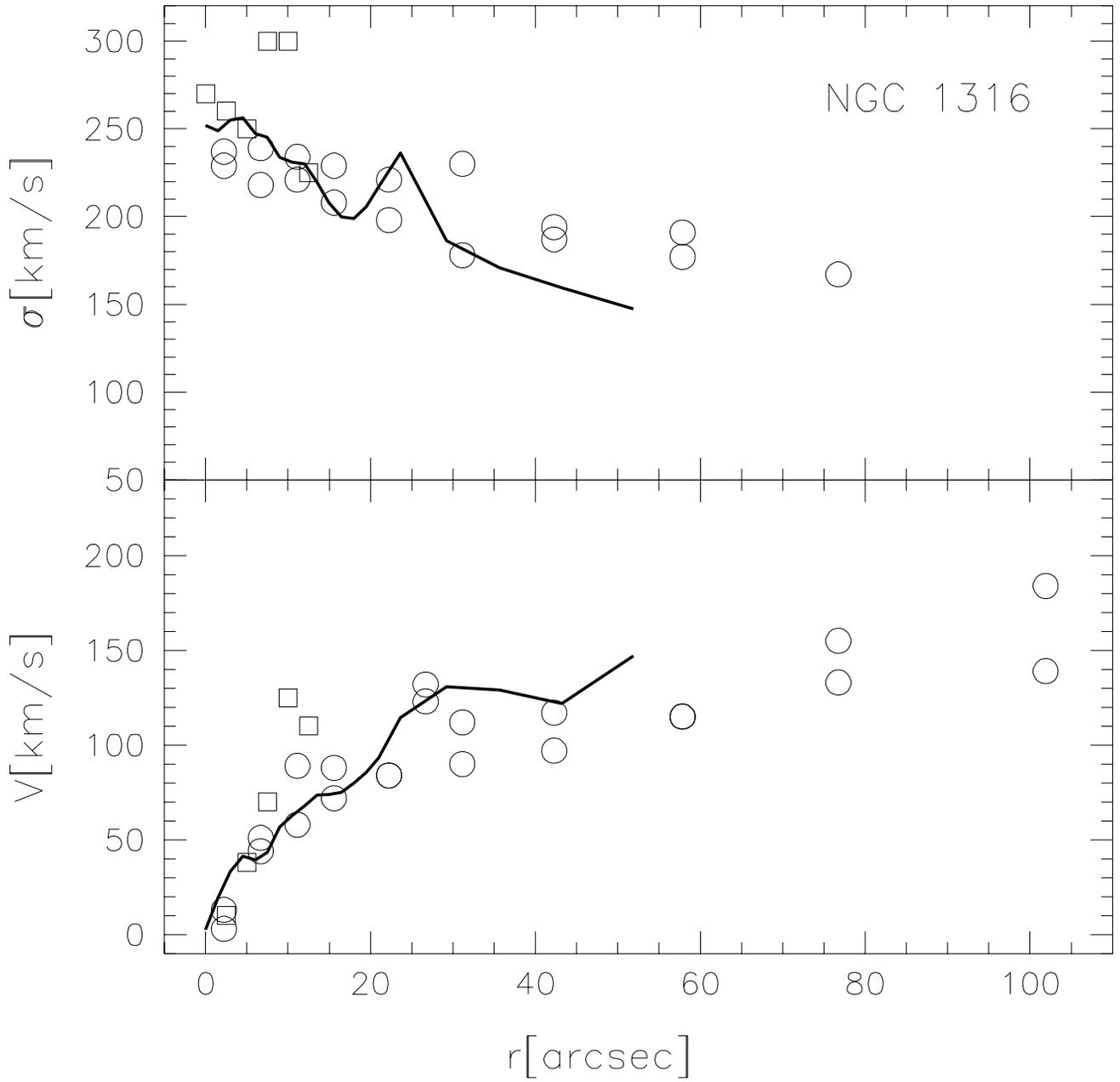

NGC 1316

Fig. 3 top

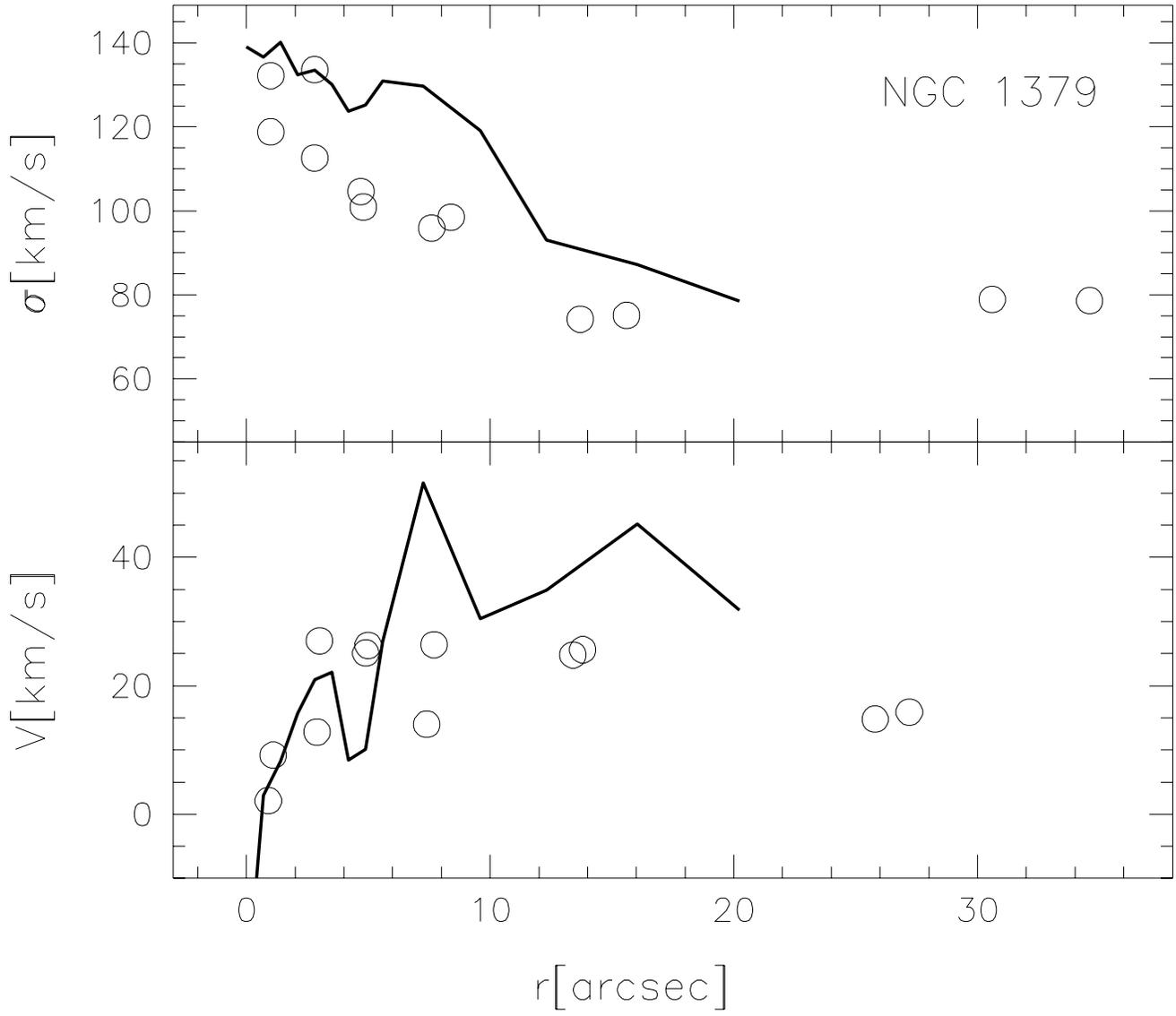

Fig. 3 center

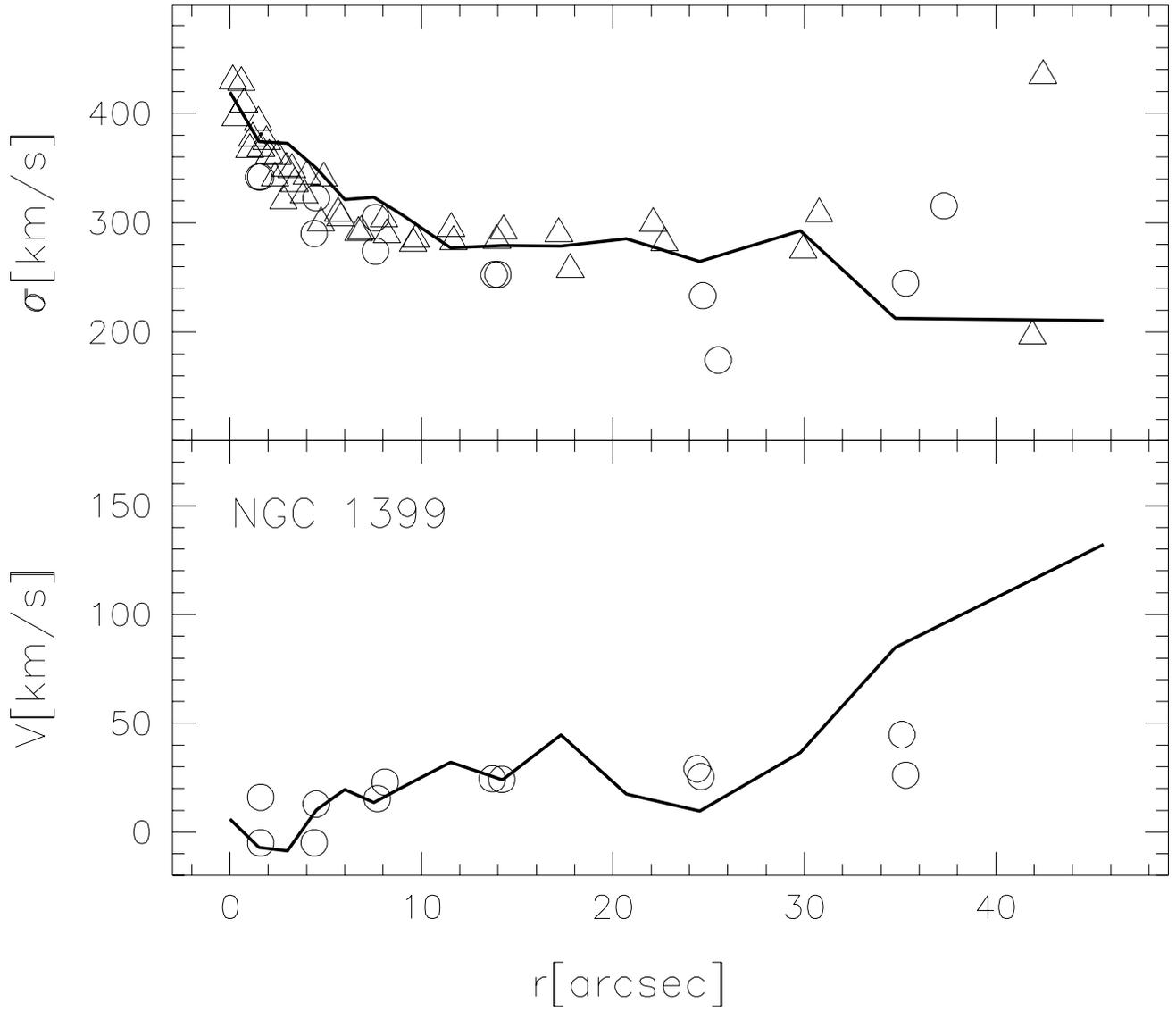

Fig. 3 bottom

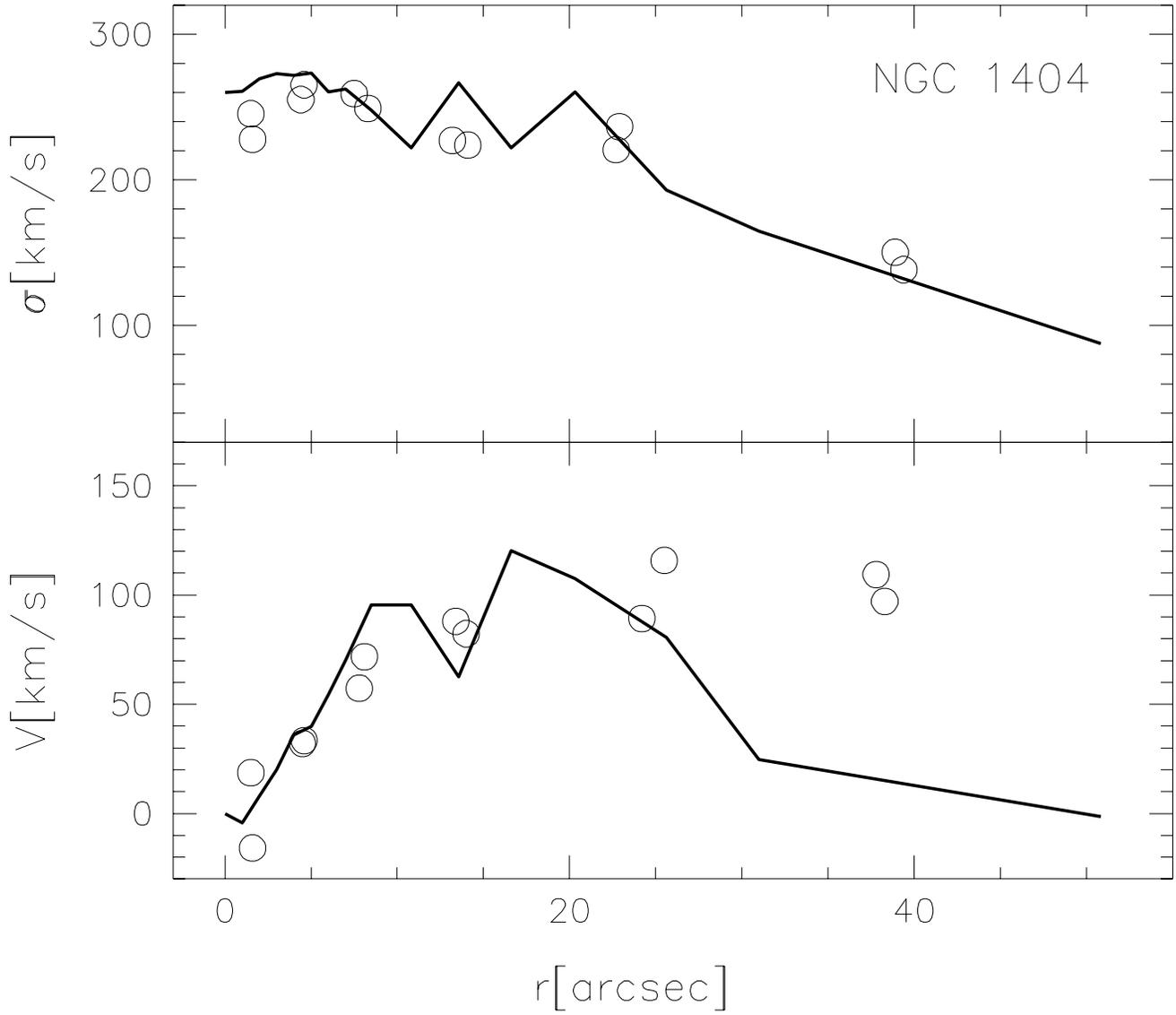

Fig. 4

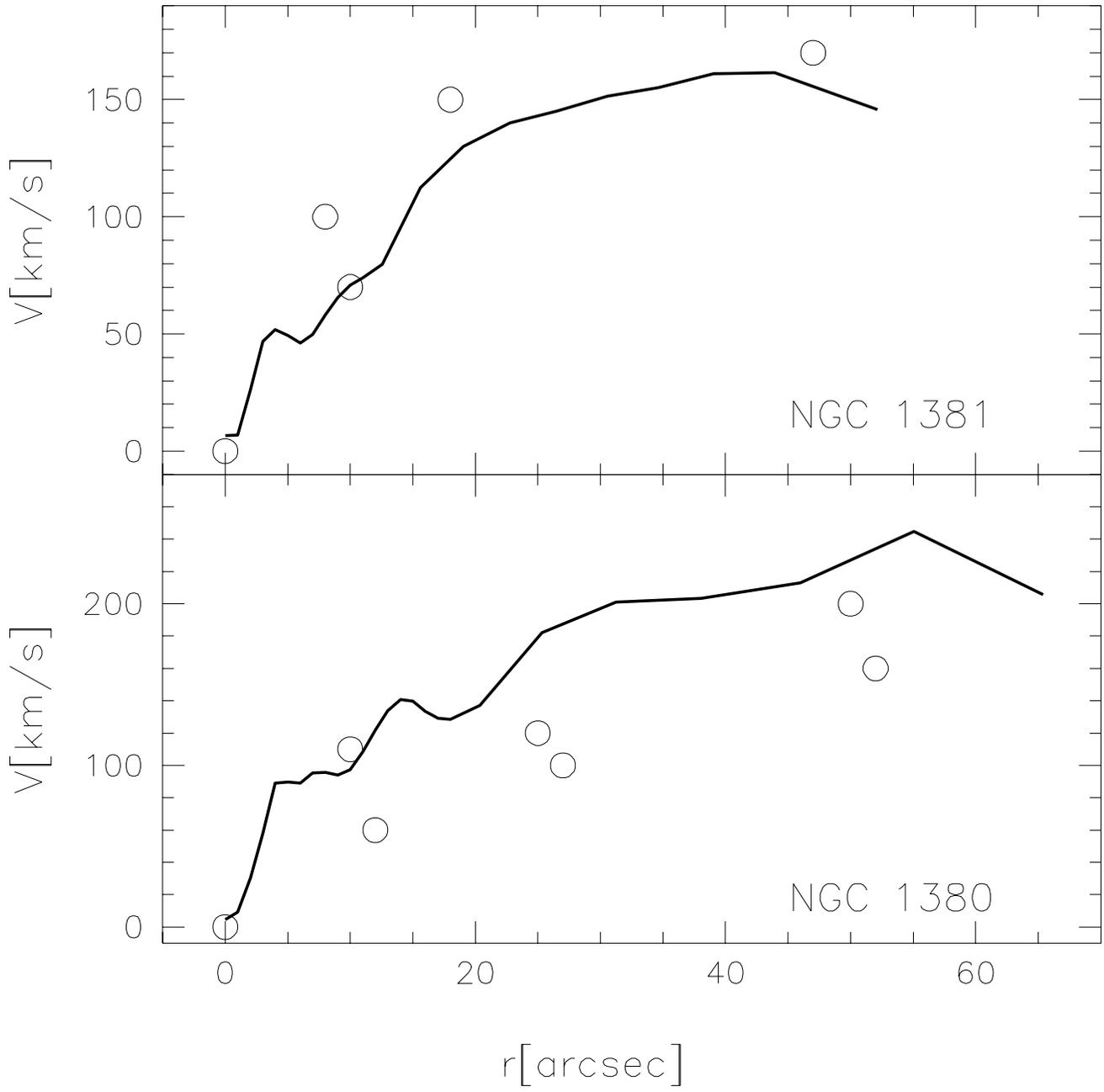

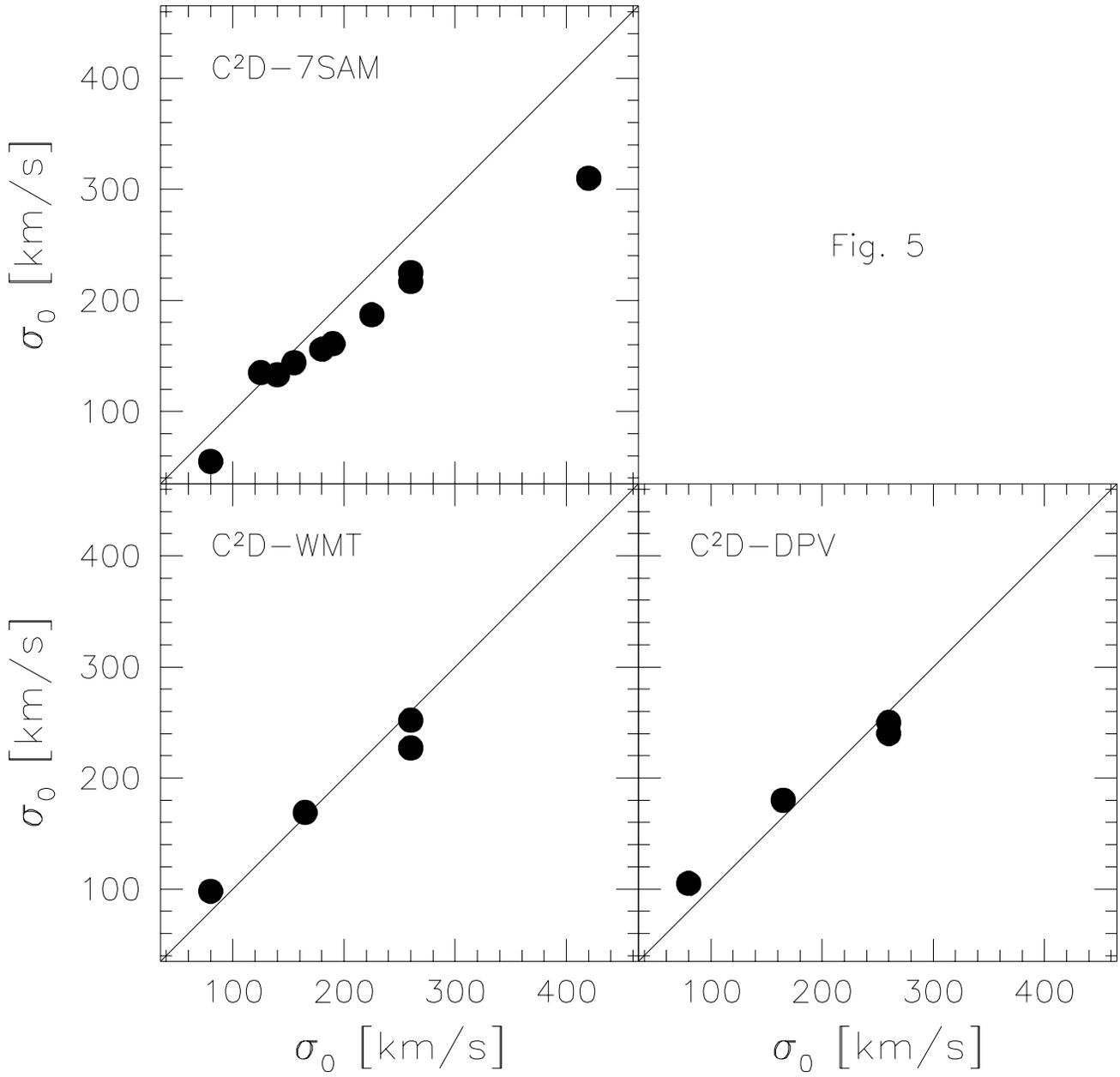

Fig. 5

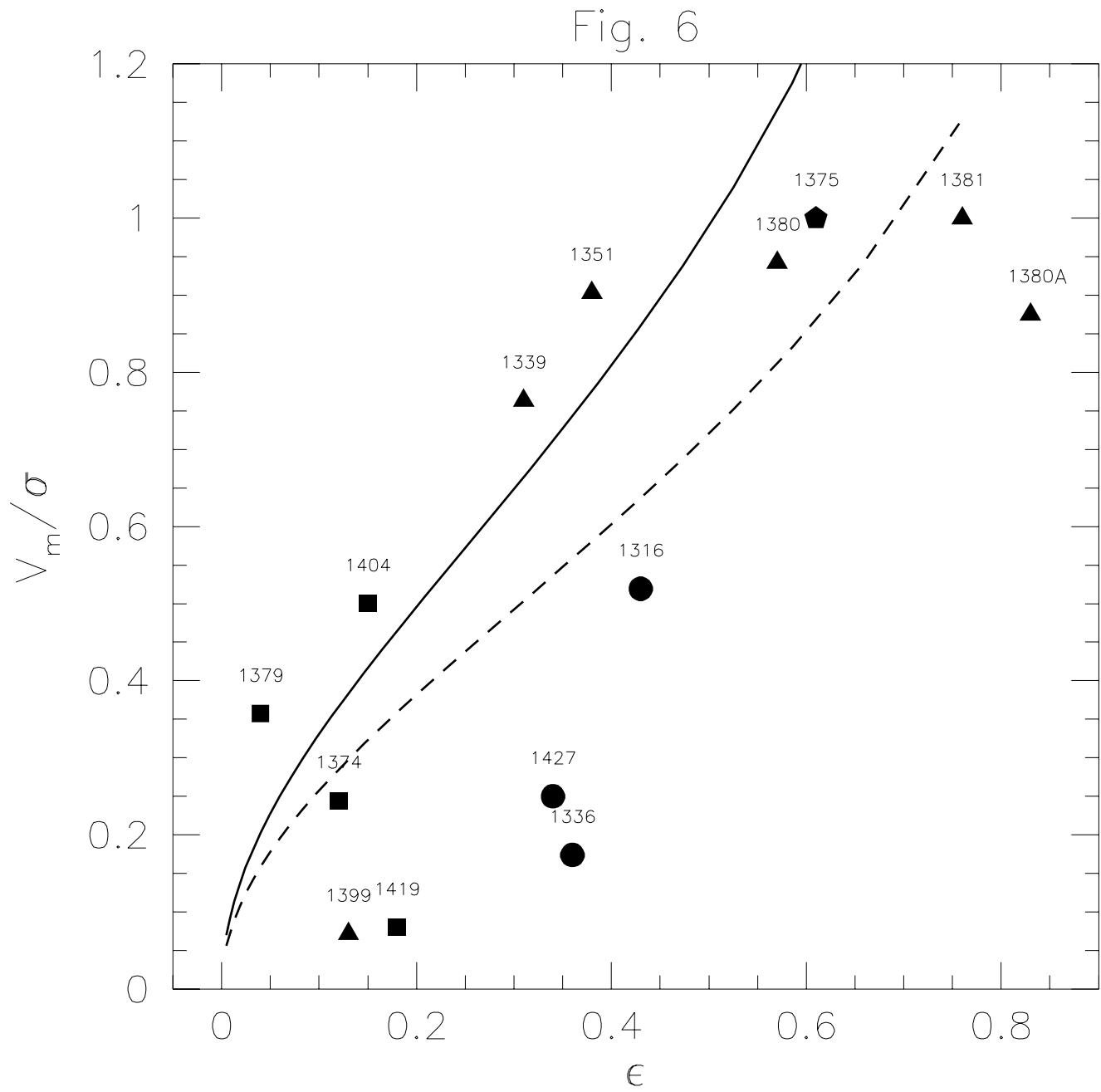

Fig. 6

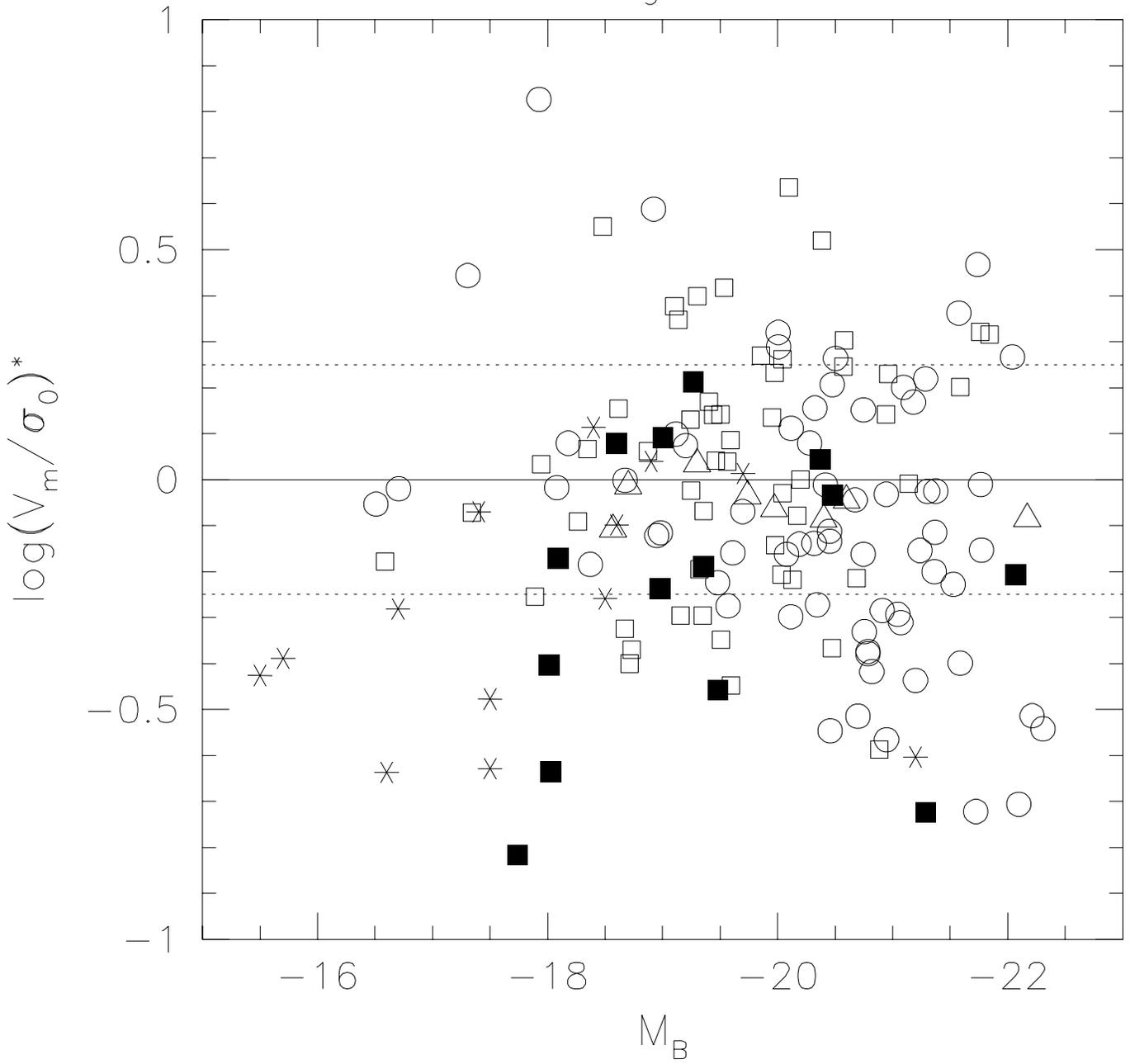

Fig. 7

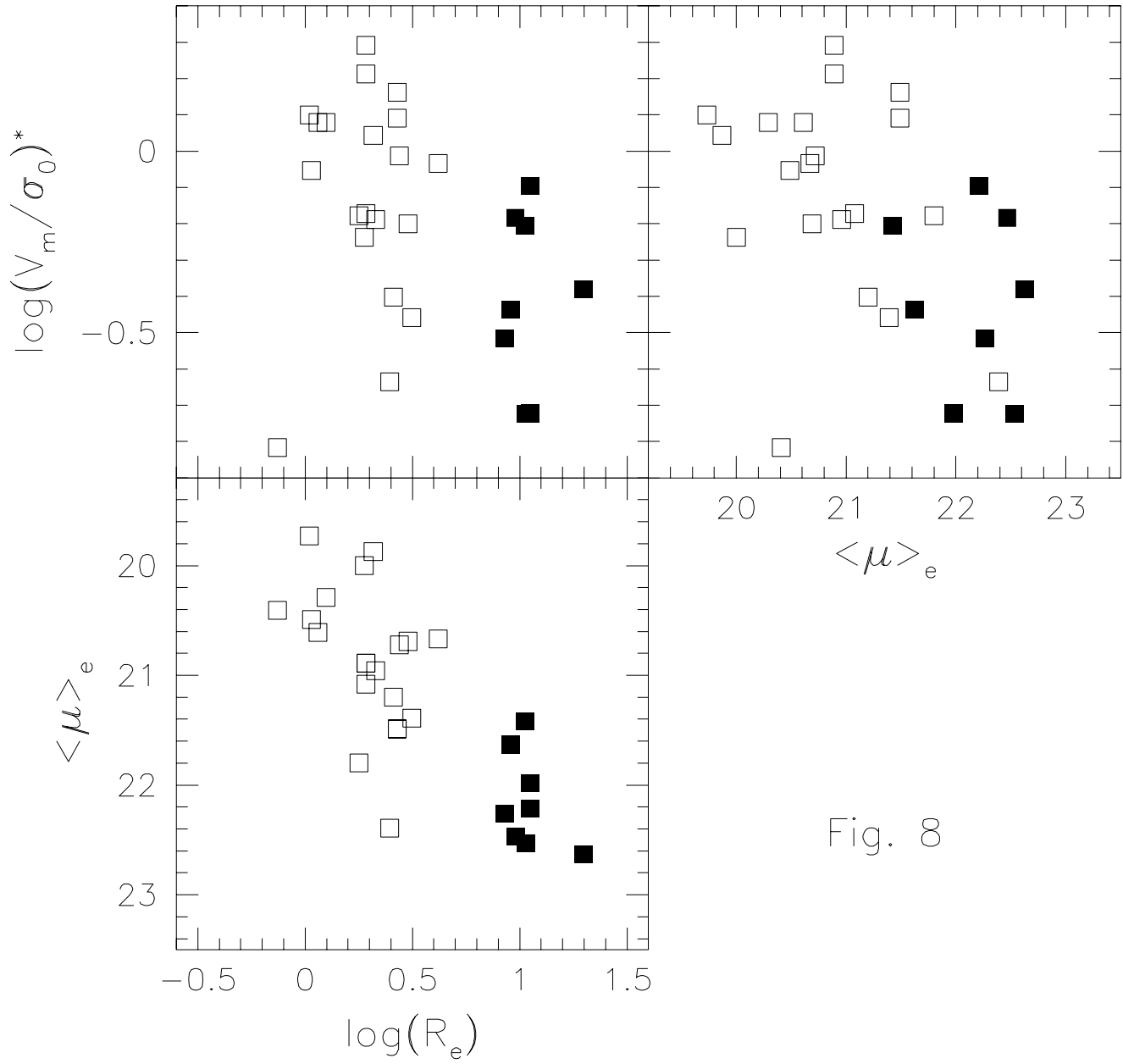

Fig. 8

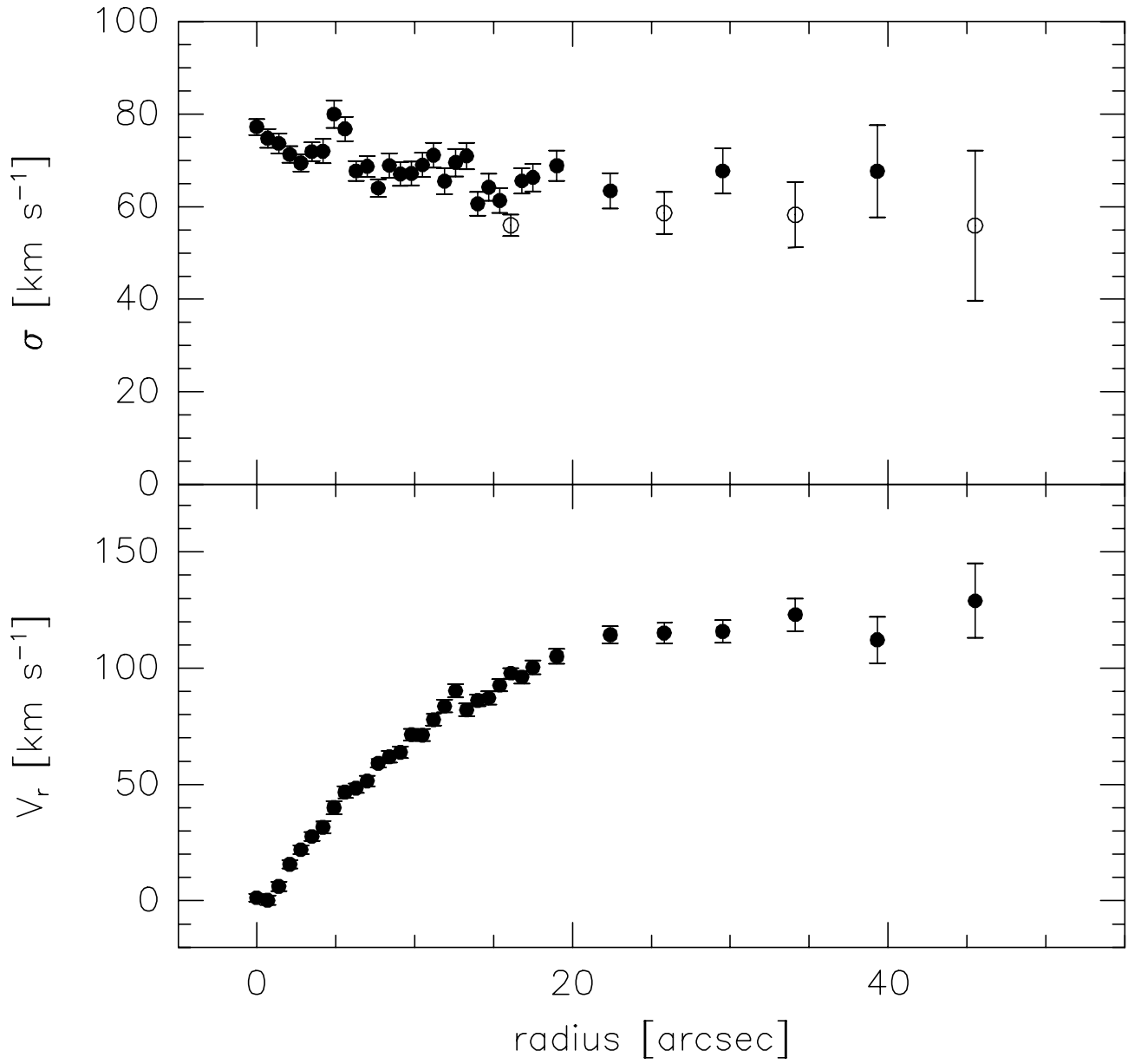

IC1963 P.A. 84°

NGC1316 P.A. 50˚

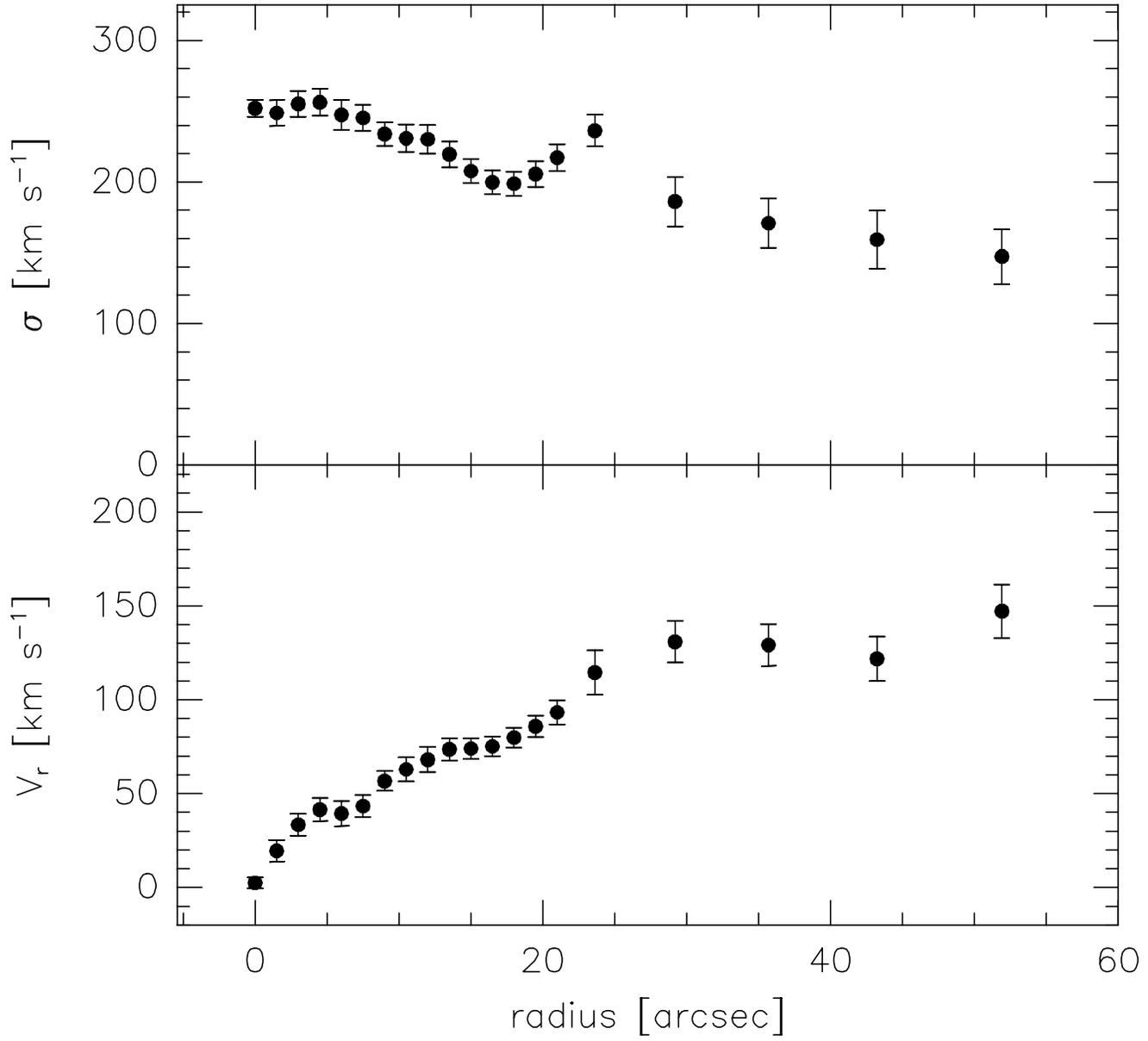

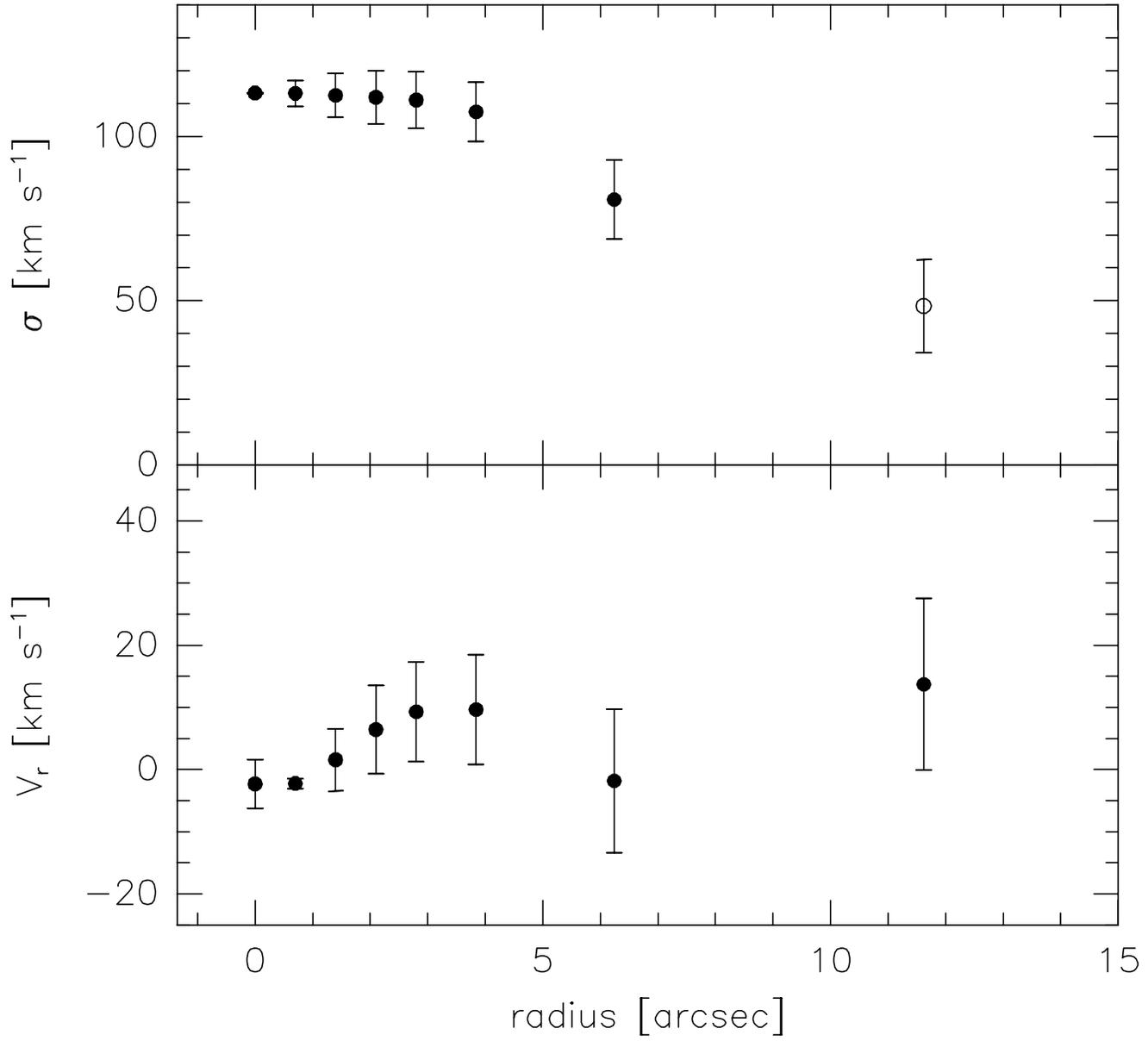

NGC1336 P.A. 20°

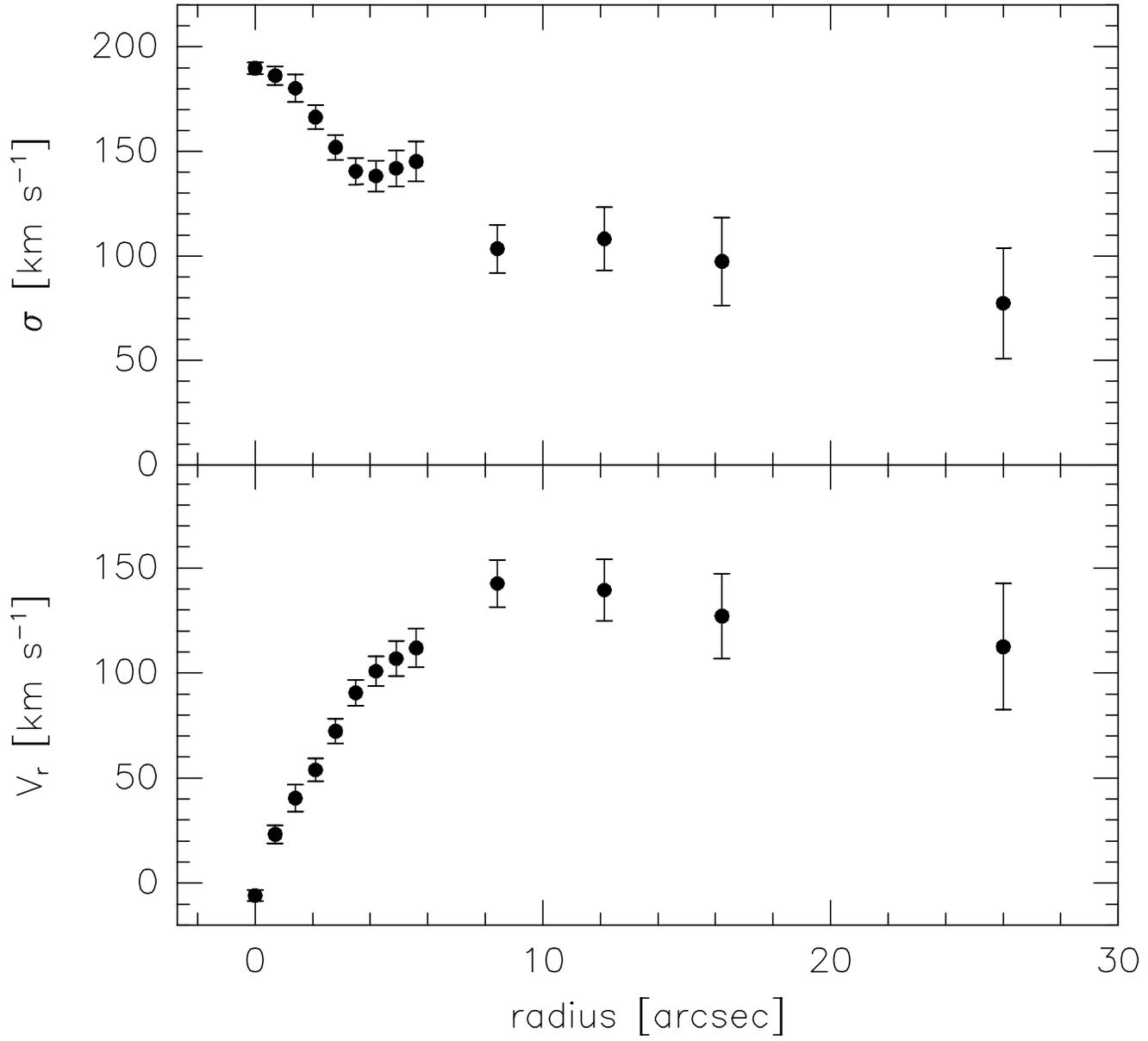

NGC1339 P.A. 175°

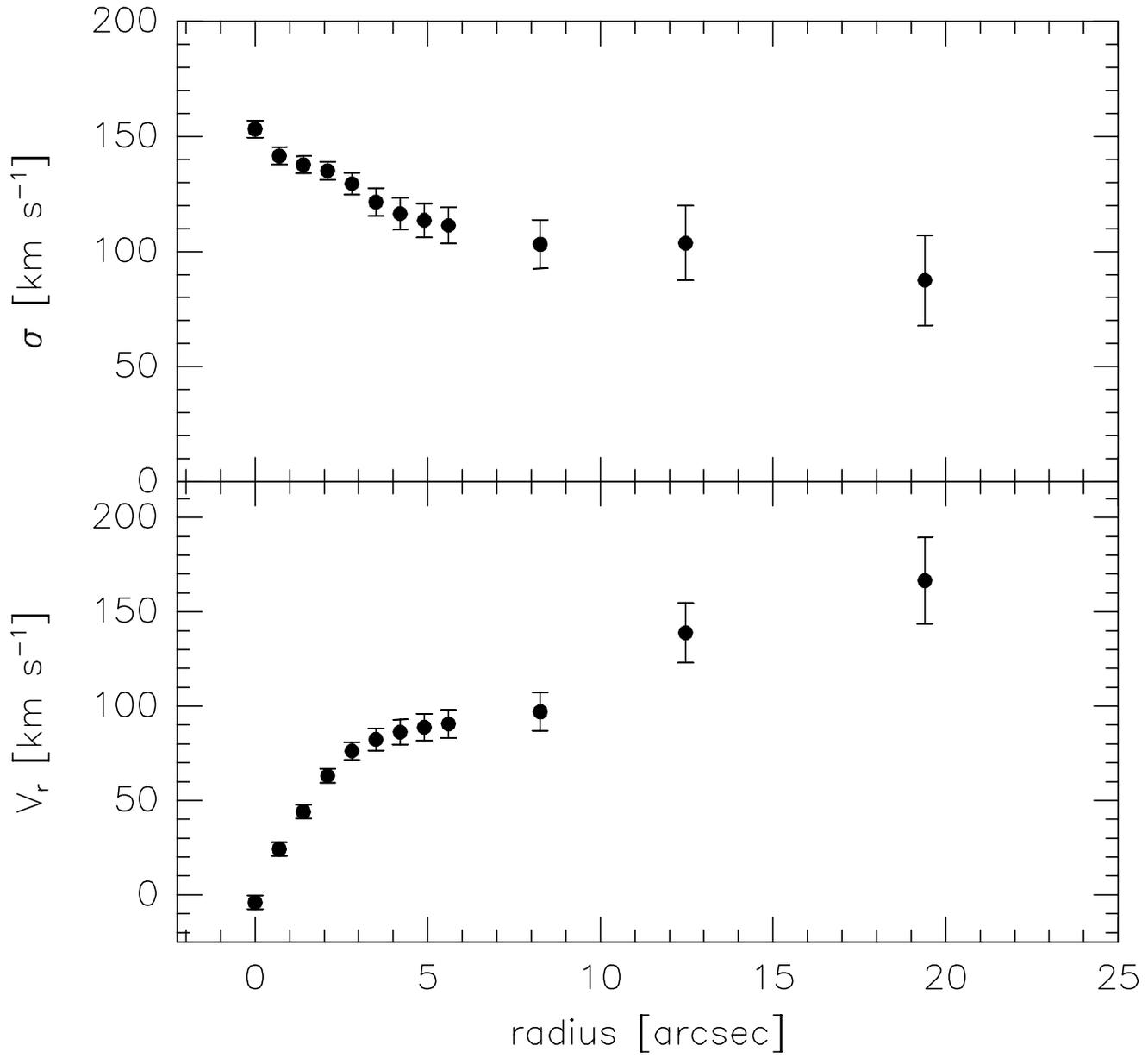

NGC1351 P.A. 141°

NGC1374 P.A. 119˚

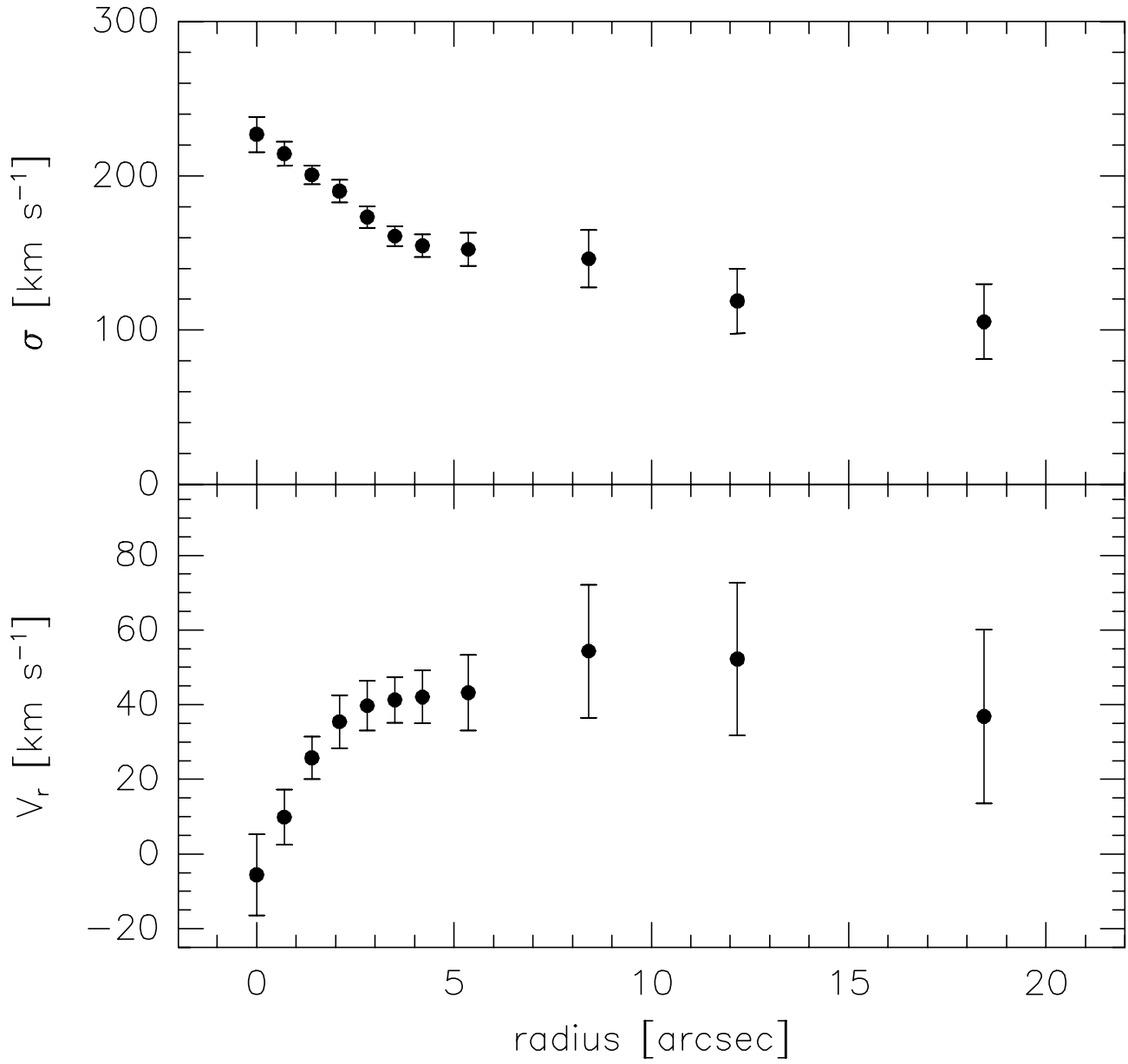

NGC1375 P.A. 88°

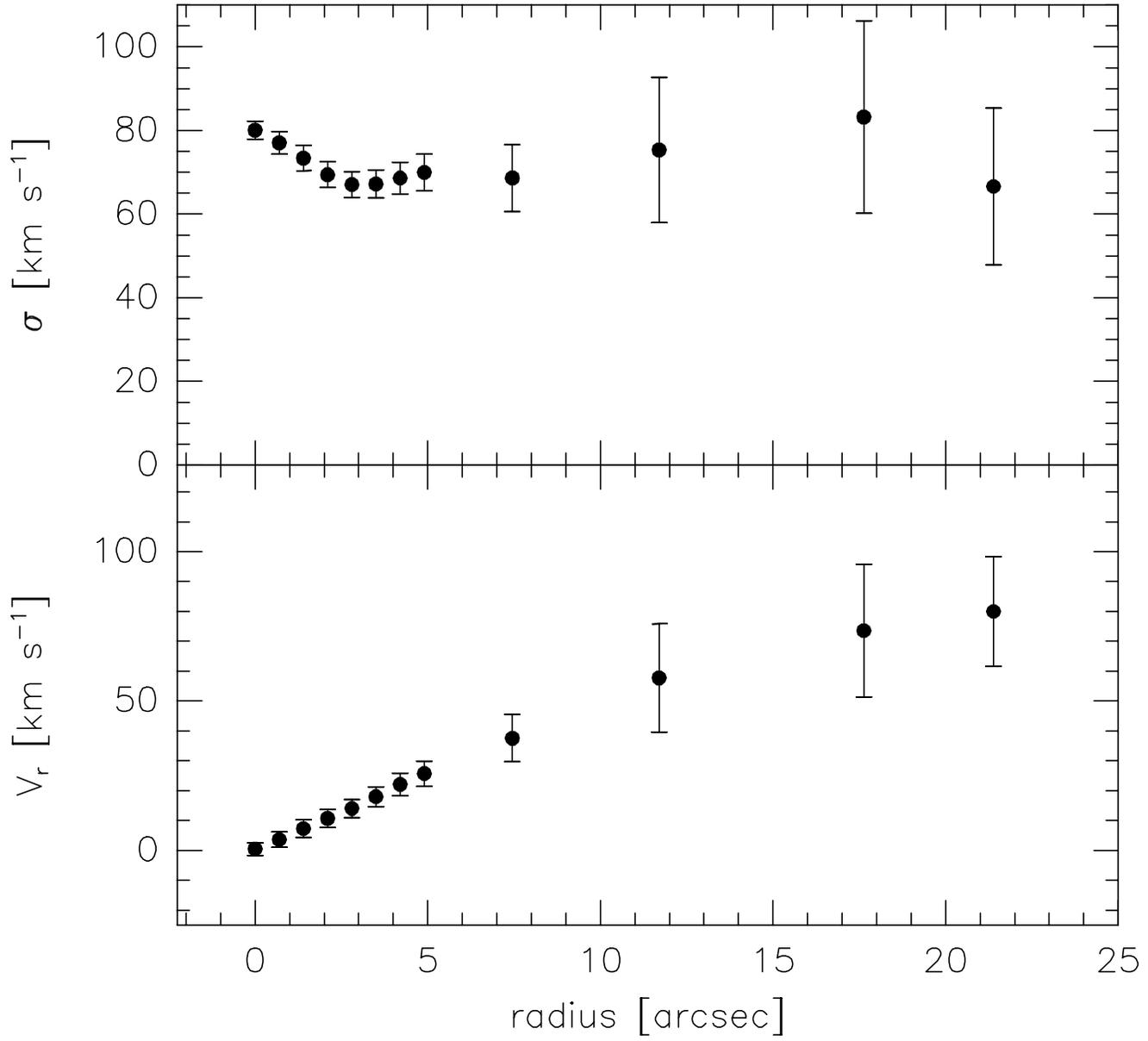

NGC1379 P.A. 7°

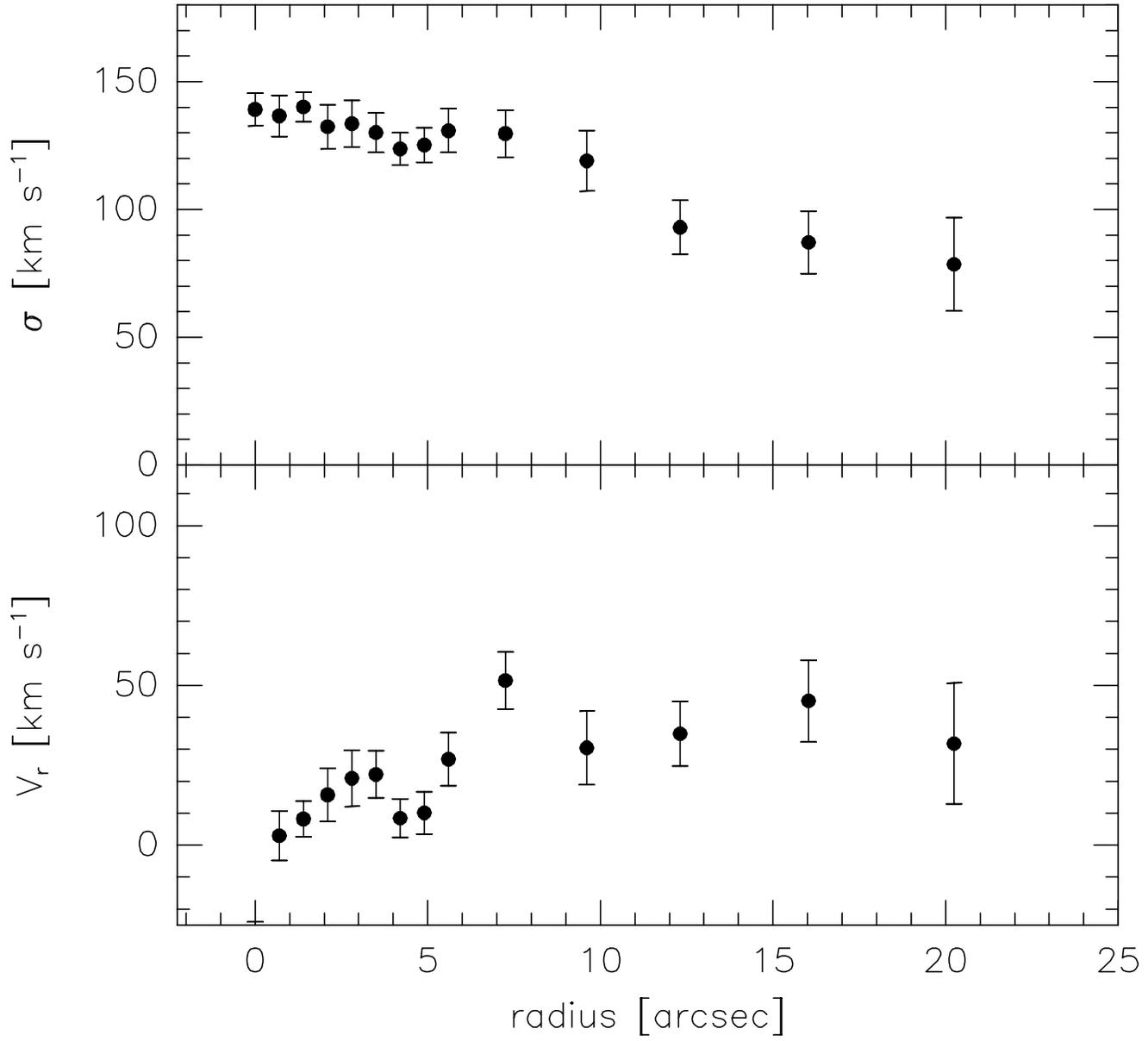

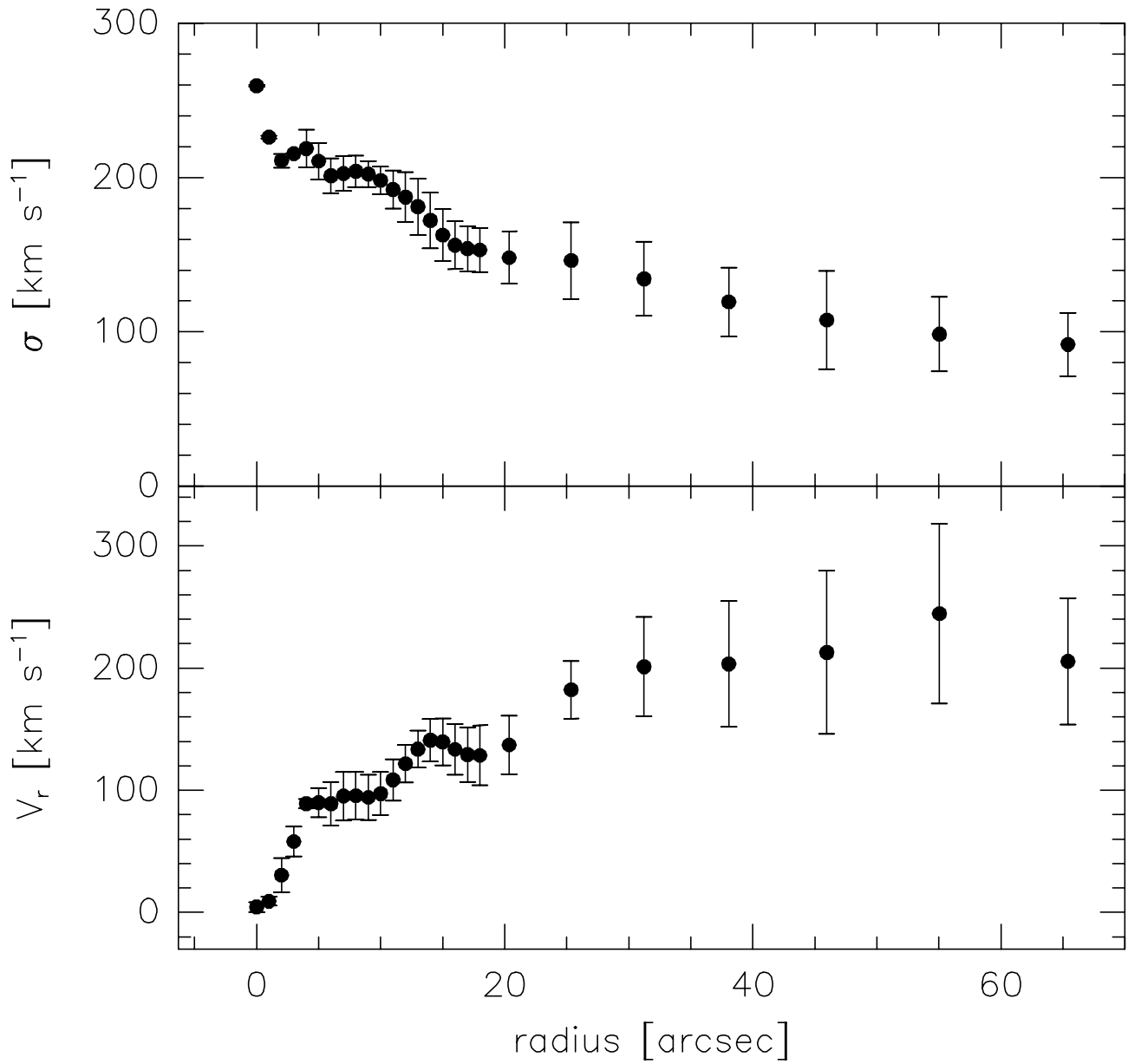

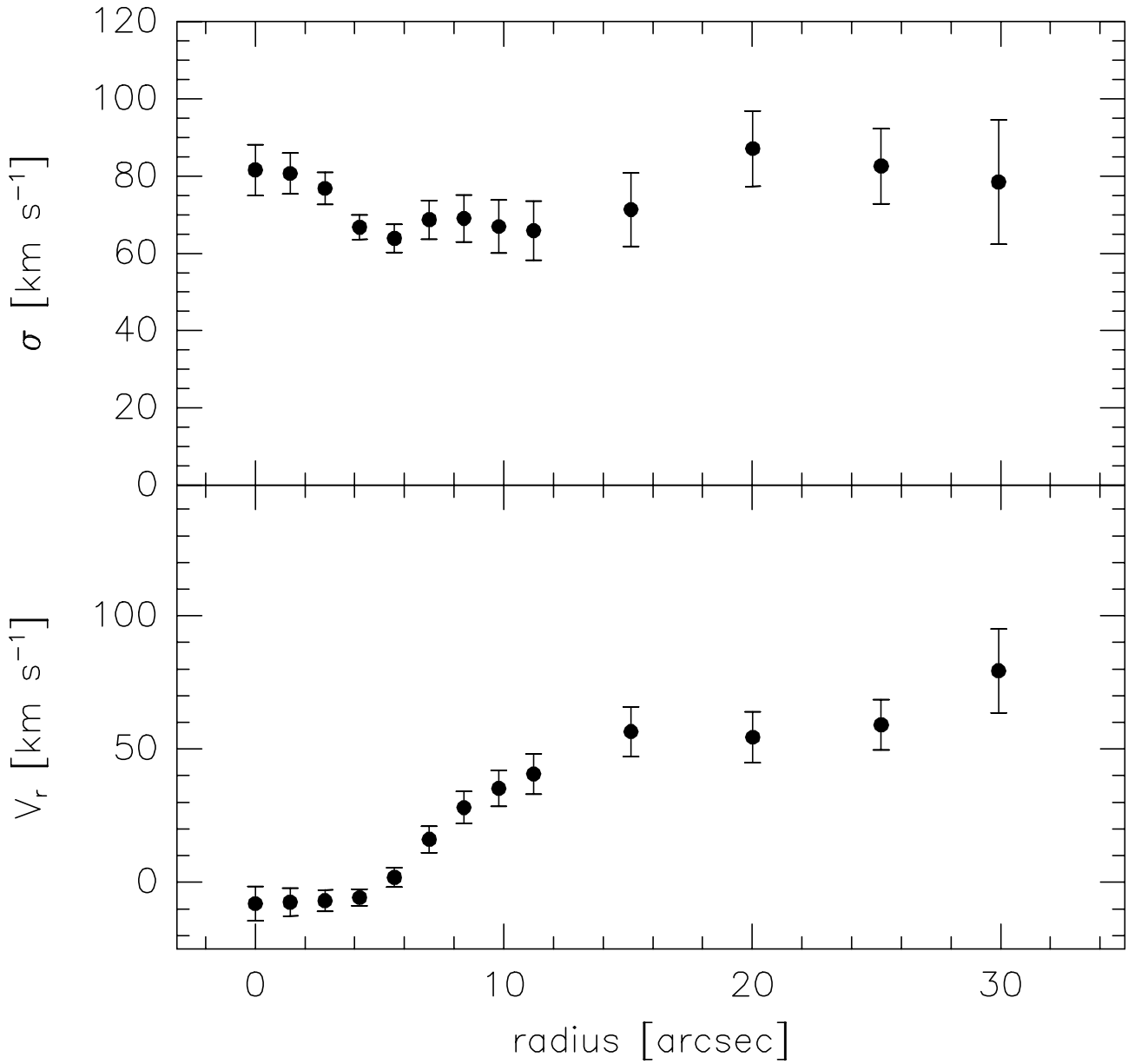

NGC1381 P.A. 139˚

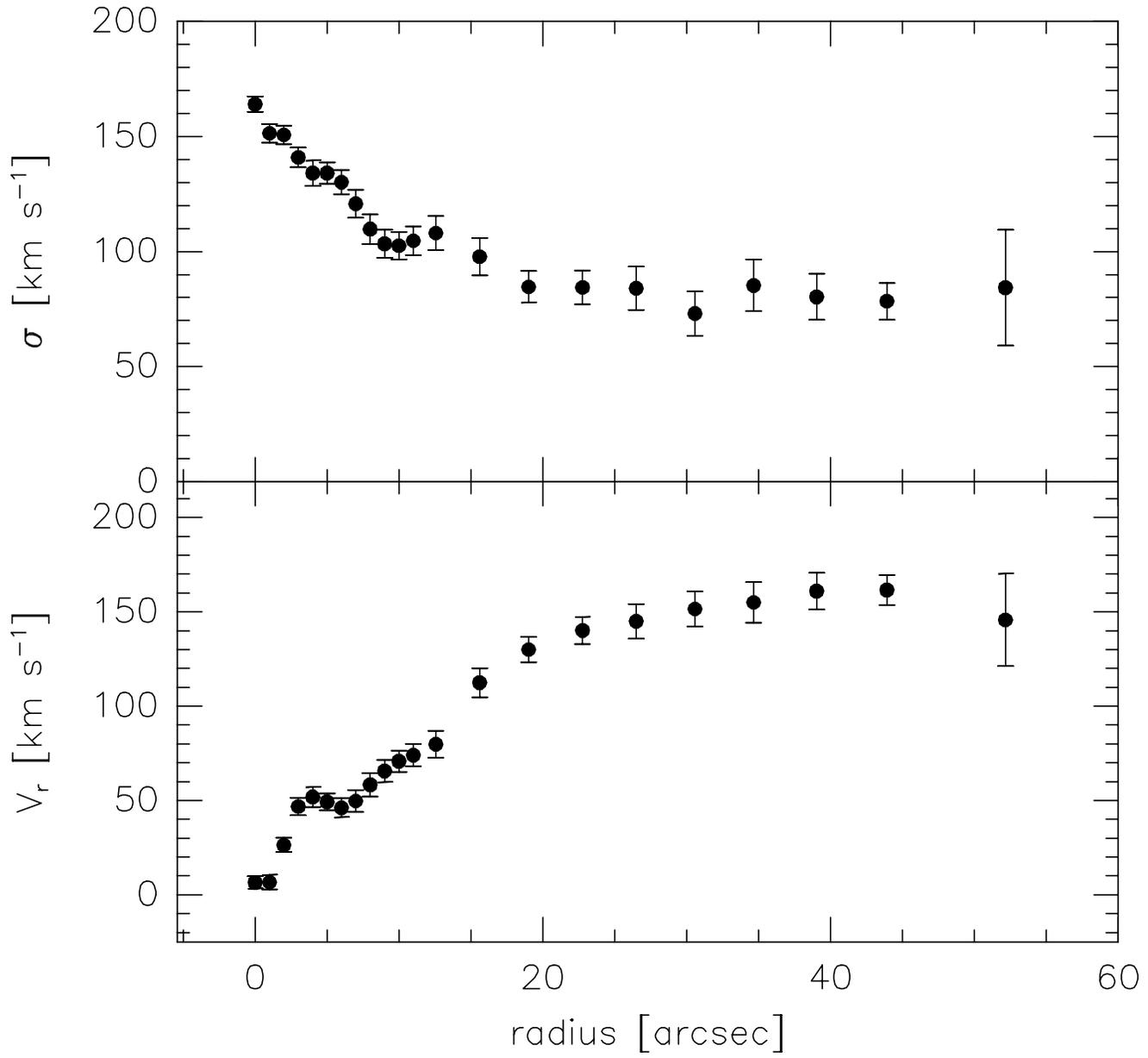

NGC1399 P.A. 112°

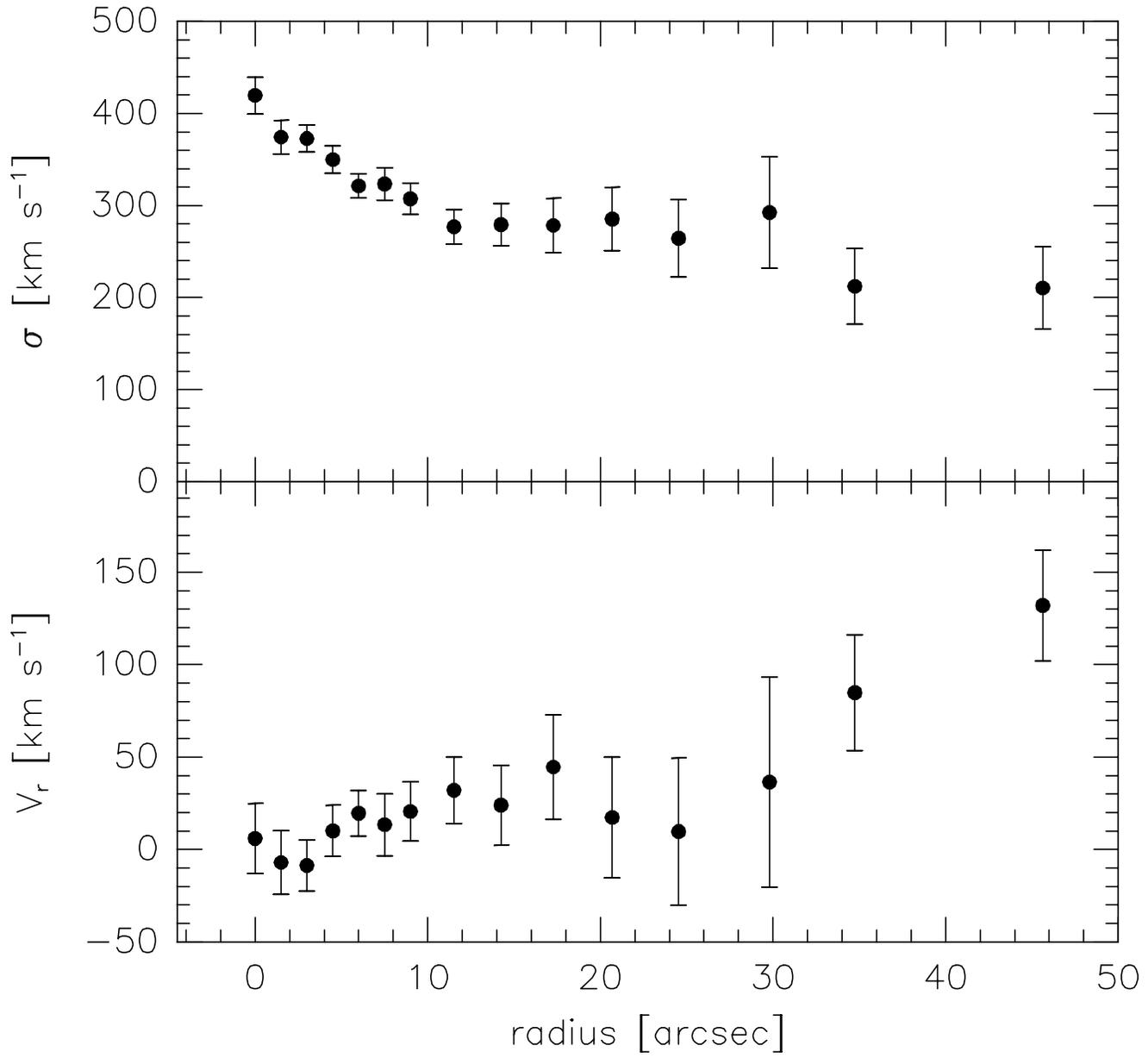

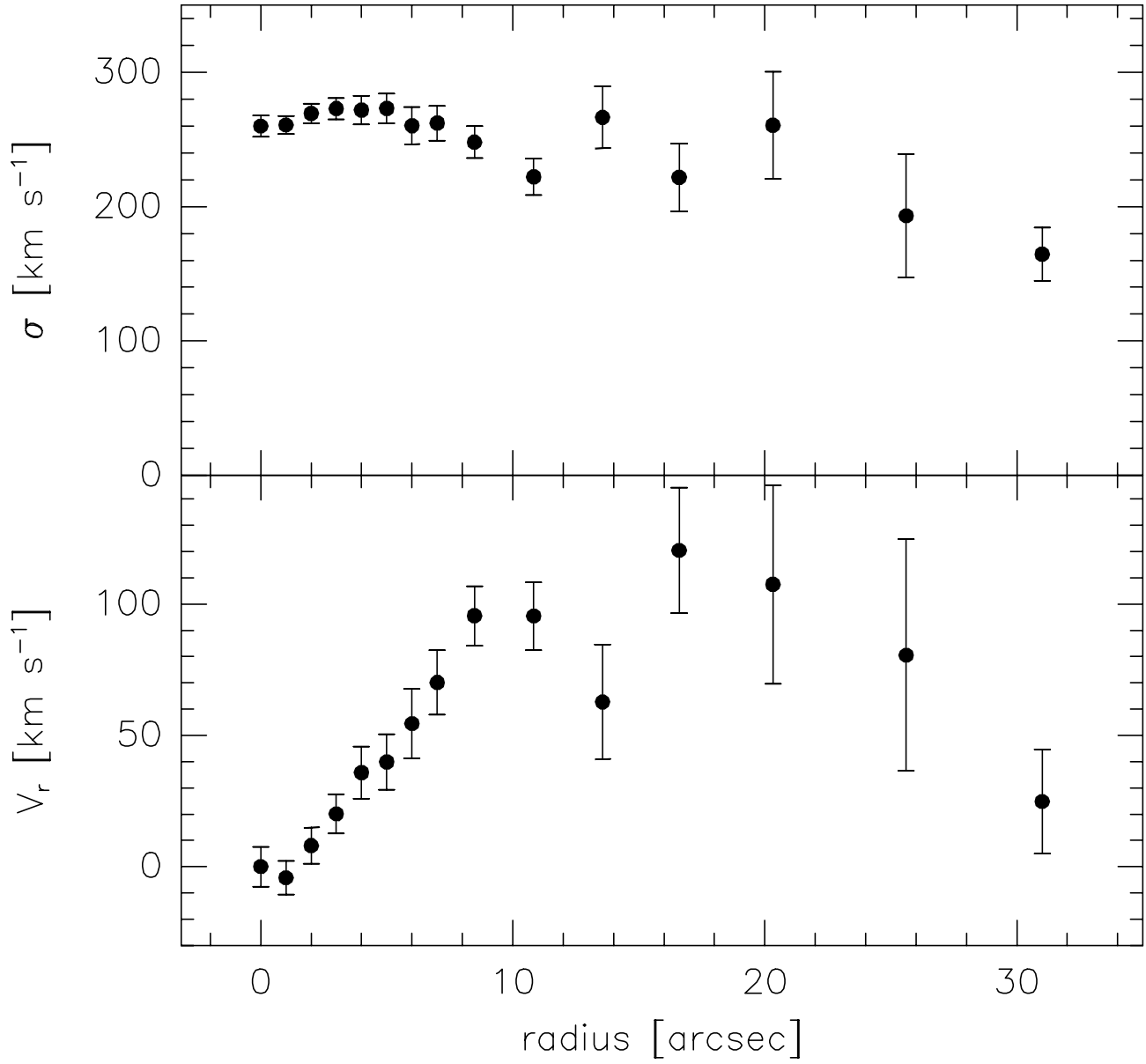

NGC1404 P.A. 159°

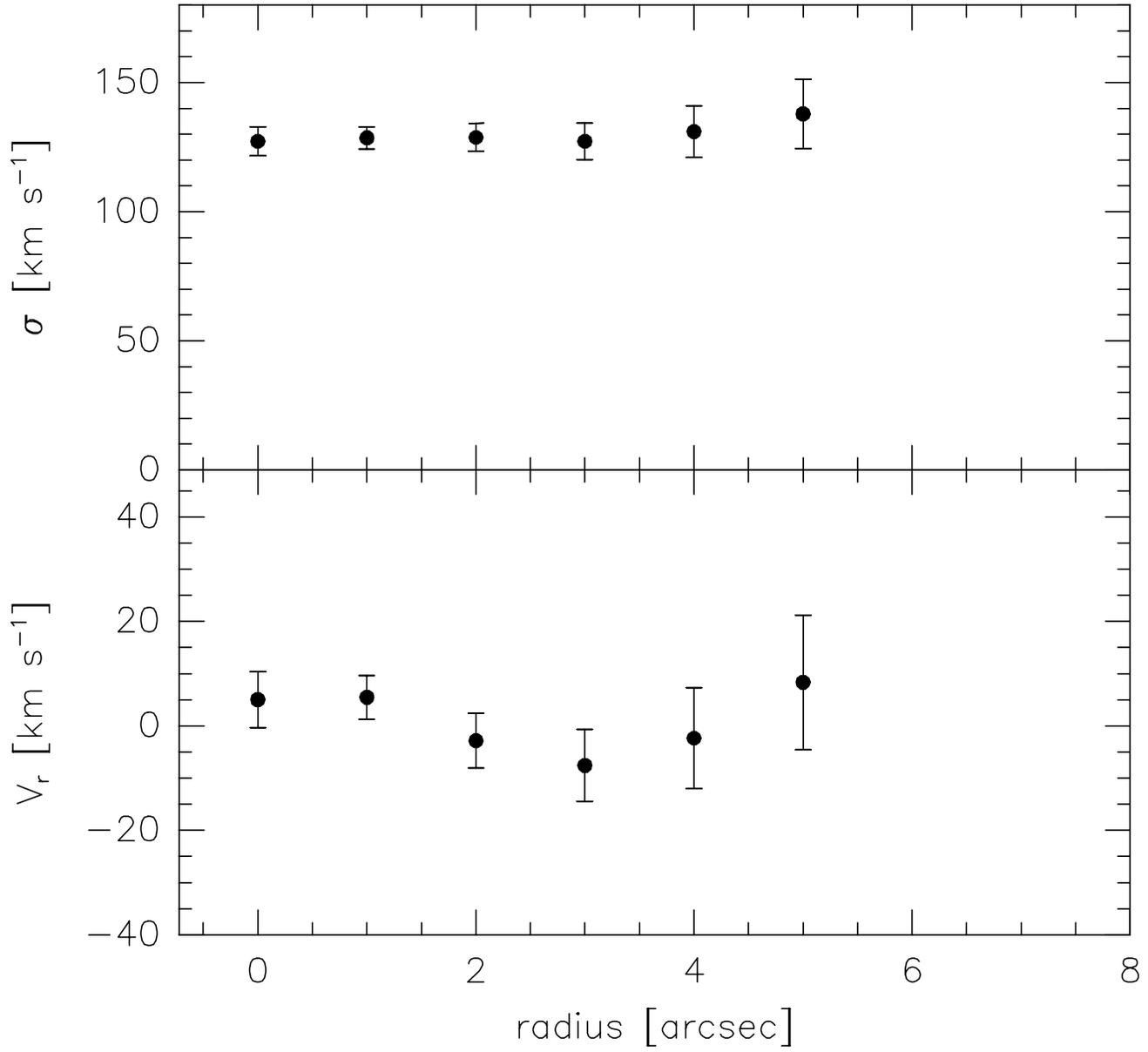

NGC1427 P.A. 79°

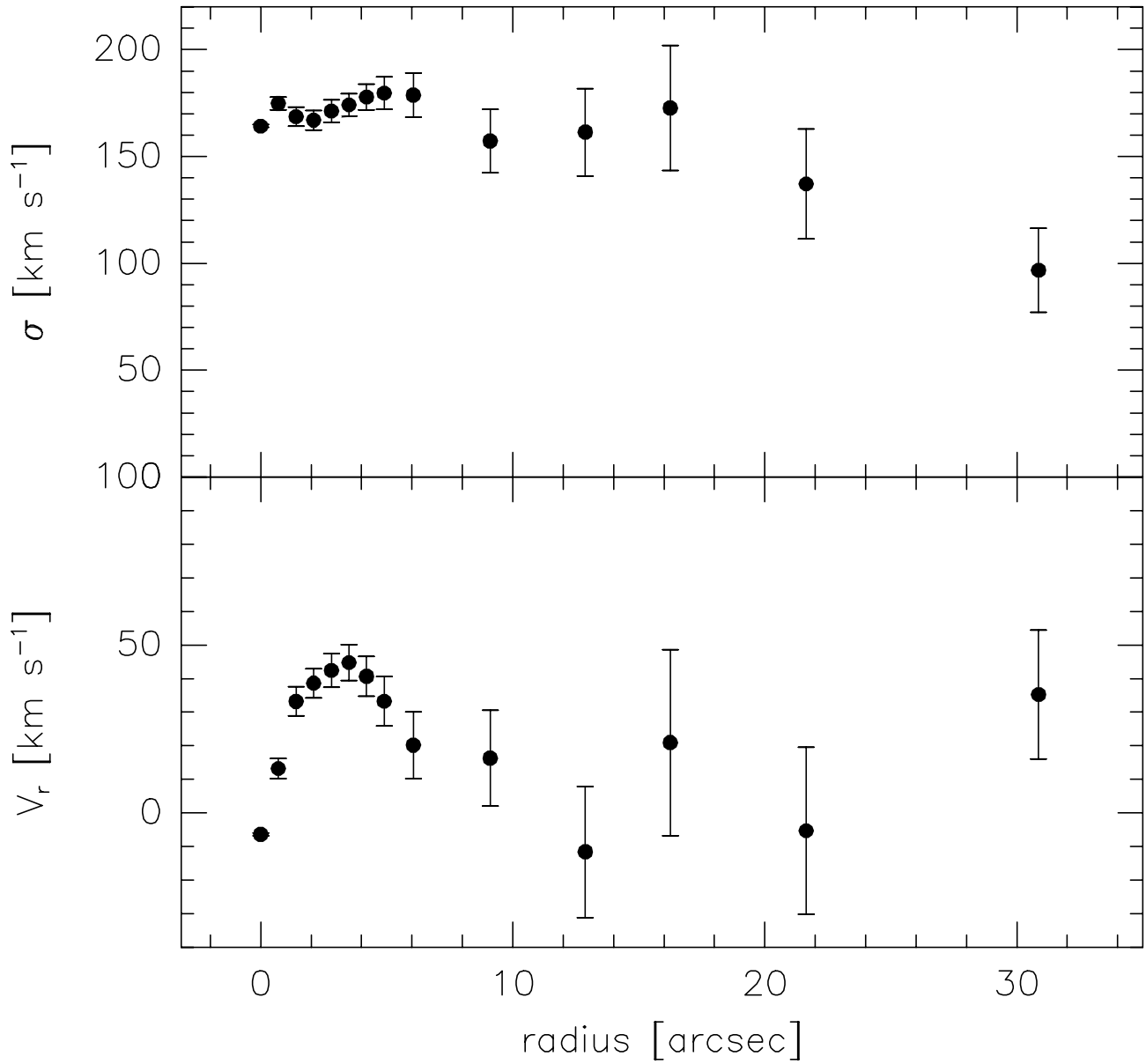